\documentclass[aps,twocolumn,superscriptaddress,floatfix,nofootinbib,showpacs,amsmath,amssymb,altaffilletter]{revtex4-1}

\input{ds.def}

\begin{document}

\title{Measurement of the liquid argon energy response to nuclear and electronic recoils}
\date{\today}

\author{P.~Agnes}\affiliation{Department of Physics, University of Houston, Houston, TX 77204, USA}\affiliation{APC, Universit\'e Paris Diderot, CNRS/IN2P3, CEA/Irfu, Obs. de Paris, Sorbonne Paris Cit\'e, Paris 75205, France}
\author{J.~Dawson}\affiliation{APC, Universit\'e Paris Diderot, CNRS/IN2P3, CEA/Irfu, Obs. de Paris, Sorbonne Paris Cit\'e, Paris 75205, France}
\author{S.~De~Cecco}\affiliation{LPNHE Paris, Sorbonne Universit\'e, Universit\'e Paris Diderot, CNRS/IN2P3, Paris 75252, France}
\author{A.~Fan}\affiliation{Department of Physics and Astronomy, University of California, Los Angeles, CA 90095, USA}
\author{G.~Fiorillo}\affiliation{Department of Physics, Universit\`a degli Studi Federico II, Napoli 80126, Italy}\affiliation{Istituto Nazionale di Fisica Nucleare, Sezione di Napoli, Napoli 80126, Italy}
\author{D.~Franco}\affiliation{APC, Universit\'e Paris Diderot, CNRS/IN2P3, CEA/Irfu, Obs. de Paris, Sorbonne Paris Cit\'e, Paris 75205, France}
\author{C.~Galbiati}\affiliation{Department of Physics, Princeton University, Princeton, NJ 08544, USA}
\author{C.~Giganti}\affiliation{LPNHE Paris, Sorbonne Universit\'e, Universit\'e Paris Diderot, CNRS/IN2P3, Paris 75252, France}
\author{T.~N.~Johnson}\email{tesjohns@ucdavis.edu}\affiliation{Department of Physics, University of California, Davis, CA 95616, USA}
\author{G.~Korga}\affiliation{Department of Physics, University of Houston, Houston, TX 77204, USA}\affiliation{Laboratori Nazionali del Gran Sasso, Assergi AQ 67010, Italy}
\author{D.~Kryn}\affiliation{APC, Universit\'e Paris Diderot, CNRS/IN2P3, CEA/Irfu, Obs. de Paris, Sorbonne Paris Cit\'e, Paris 75205, France}
\author{M.~Lebois}\affiliation{Institut de Physique Nuclaire Orsay, F91406 Orsay, France}
\author{A.~Mandarano}\affiliation{Gran Sasso Science Institute, L'Aquila 67100, Italy}\affiliation{Laboratori Nazionali del Gran Sasso, Assergi AQ 67010, Italy}
\author{C.~J.~Martoff}\affiliation{Department of Physics, Temple University, Philadelphia, PA 19122, USA}
\author{A.~Navrer-Agasson~}\affiliation{LPNHE Paris, Sorbonne Universit\'e, Universit\'e Paris Diderot, CNRS/IN2P3, Paris 75252, France}
\author{E.~Pantic}\affiliation{Department of Physics, University of California, Davis, CA 95616, USA}
\author{L.~Qi}\affiliation{Institut de Physique Nuclaire Orsay, F91406 Orsay, France}
\author{A.~Razeto}\affiliation{Laboratori Nazionali del Gran Sasso, Assergi AQ 67010, Italy}
\author{A.~L.~Renshaw}\affiliation{Department of Physics, University of Houston, Houston, TX 77204, USA}
\author{Q.~Riffard}\email{riffard@apc.in2p3.fr}\affiliation{APC, Universit\'e Paris Diderot, CNRS/IN2P3, CEA/Irfu, Obs. de Paris, Sorbonne Paris Cit\'e, Paris 75205, France}
\author{B.~Rossi}\affiliation{Istituto Nazionale di Fisica Nucleare, Sezione di Napoli, Napoli 80126, Italy}
\author{C.~Savarese}\affiliation{Gran Sasso Science Institute, L'Aquila 67100, Italy}\affiliation{Laboratori Nazionali del Gran Sasso, Assergi AQ 67010, Italy}
\author{B. Schlitzer}\affiliation{Department of Physics, University of California, Davis, CA 95616, USA}
\author{Y.~Suvorov}\affiliation{Department of Physics and Astronomy, University of California, Los Angeles, CA 90095, USA}\affiliation{Laboratori Nazionali del Gran Sasso, Assergi AQ 67010, Italy}
\author{A.~Tonazzo}\affiliation{APC, Universit\'e Paris Diderot, CNRS/IN2P3, CEA/Irfu, Obs. de Paris, Sorbonne Paris Cit\'e, Paris 75205, France}
\author{H.~Wang}\affiliation{Department of Physics and Astronomy, University of California, Los Angeles, CA 90095, USA}
\author{Y.~Wang}\affiliation{Department of Physics and Astronomy, University of California, Los Angeles, CA 90095, USA}
\author{A.~W.~Watson}\affiliation{Department of Physics, Temple University, Philadelphia, PA 19122, USA}
\author{J.~N.~Wilson}\affiliation{Institut de Physique Nuclaire Orsay, F91406 Orsay, France}

\collaboration{The ARIS Collaboration}\noaffiliation

\begin{abstract}
A liquid argon time projection chamber, constructed for the Argon Response to Ionization and Scintillation (ARIS) experiment, has been exposed to the highly collimated and quasi-monoenergetic LICORNE neutron beam at the Institute de Physique Nuclaire Orsay in order to study the scintillation response to nuclear and electronic recoils. An array of liquid scintillator detectors, arranged around the apparatus, tag scattered neutrons and select nuclear recoil energies in the  [7, 120]~keV energy range.
The relative scintillation efficiency of nuclear recoils was measured to high precision at null field, and the ion-electron recombination probability was extracted for  a range of applied electric fields.
Single-scattered Compton electrons, produced by gammas emitted from the de-excitation of \Lim in coincidence with the beam pulse, along with calibration gamma sources,  are used to extract the recombination probability as a function of energy and electron drift field. The ARIS results have been compared with three recombination probability parameterizations (Thomas-Imel, Doke-Birks, and PARIS), allowing for the definition of a fully comprehensive model of the liquid argon response to nuclear and electronic recoils  down to a few keV range.  The constraints provided by ARIS to the liquid argon response at low energy  allow the reduction of systematics affecting the sensitivity of dark matter search experiments based on  liquid argon.

\end{abstract}
\maketitle

\section{Introduction}
\label{sect:intro}

The field of direct dark matter searches has experienced a significant expansion in the past decade, with a growing number of experiments striving to increase the sensitivity to signals from dark matter particles.
Direct dark matter search experiments seek a possible interaction between dark matter and Standard Model matter in specialized, low-background detectors deployed in underground laboratories. 
The absence of an unambiguous observation of Weakly Interacting Massive Particle (WIMP) signals in recent years has pushed experiments to increase their sensitivity by simultaneously reducing the background, enlarging the active detector volume, and lowering the energy threshold of the searches. 

In this context, noble liquids are ideal candidates as target materials: they are relatively inexpensive, intrinsically more pure than other materials, and  scalable to masses in the multi-ton range. Further, they are excellent scintillators ($\sim$40,000 photons/MeV) and good ionizers (10-30 eV ionization energy) in response to the passage of radiation.  

Dual-phase noble liquid time projection chamber (TPC) detectors are currently the most sensitive detectors in searches for multi-GeV mass WIMPs ~\cite{Aprile:2017tt,Akerib:2017kg,Tan:2016fv}. The detection mechanism relies on the delayed coincidence between scintillation and ionization signals generated by the passage of an interacting particle. 
The prompt scintillation light (S1 signal) is produced by the decay of excited dimers of noble atoms, which are formed after one atom is excited.  
Interactions also produce ionization electrons, drifted by an electric field toward a gaseous region, where they produce a delayed light pulse by electroluminescence (S2 signal).  
A fraction of ionization electrons, however, recombine with ions to form excited dimers which contribute to S1 and deplete the S2 signal.

With respect to other noble liquid targets, liquid argon (LAr) exhibits a powerful rejection of electronic recoil backgrounds ($>$10$^8$ discrimination power \cite{Amaudruz:2016qqa}) through the temporal pulse shape of the scintillation signal.  
The combination of this pulse-shape discrimination technique and the use of argon extracted from deep underground, highly depleted in cosmogenic isotopes~\cite{Agnes:2016fz}, makes liquid argon an ideal target for multi-ton detectors.

The sensitivity of liquid argon detectors can be enhanced by constraining the parameters of the liquid argon response to interacting particles, such as the quenching of nuclear recoils and the electron-ion recombination effect. These parameters are difficult to constrain in large detectors with external sources, because of the passive materials which suppress interactions in the target.  Alternatively, the liquid argon response can be measured by auxiliary calibration experiments which exploit small-scale detector setups exposed to neutron and gamma beams.  These experiments, tailored specifically for measurements of the liquid argon response, are able to accurately explore the low energy ranges for nuclear and electronic recoils under controlled conditions.

\begin{figure}[t]
\includegraphics[width=0.99\columnwidth]{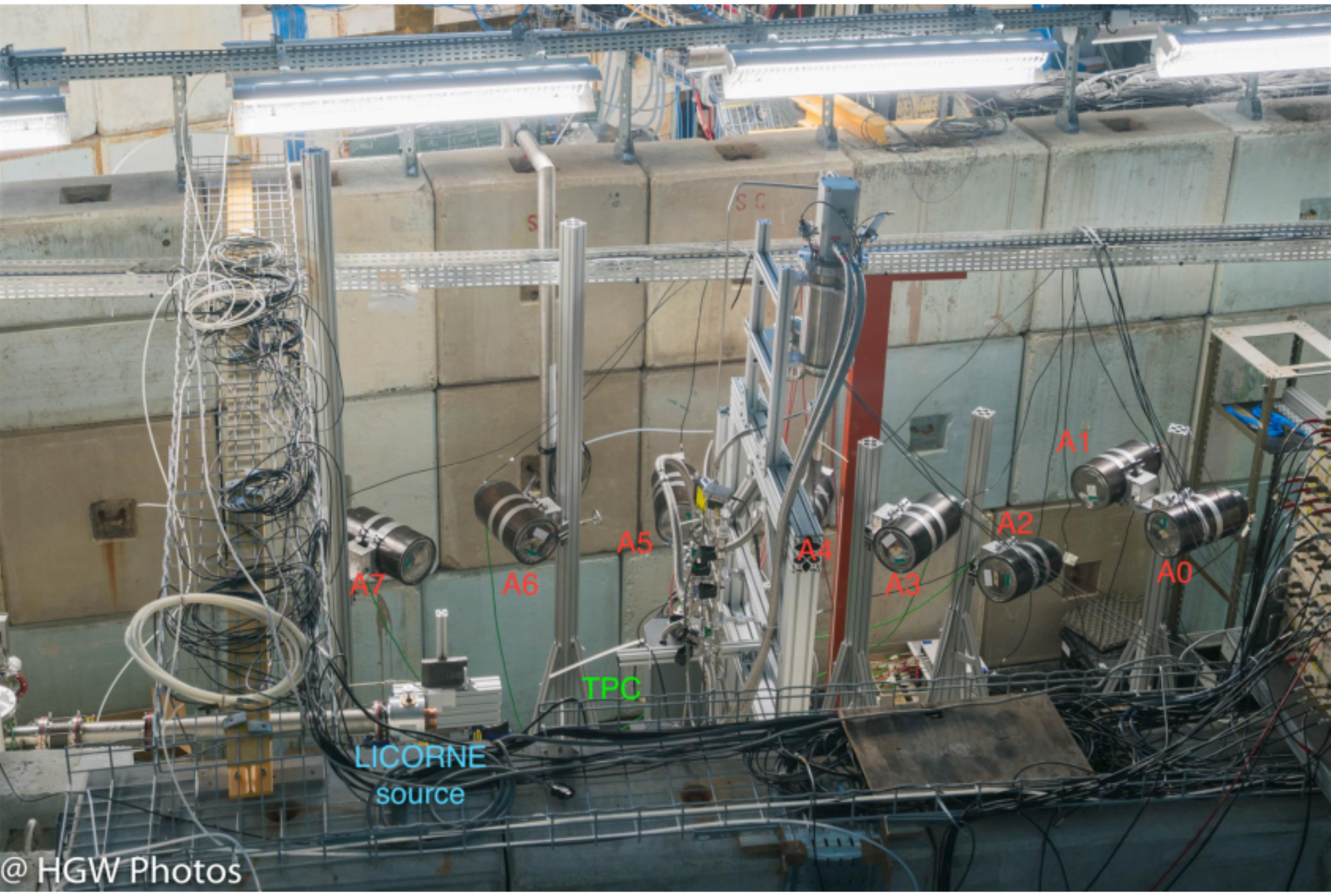}
\caption{Picture of the ARIS setup in the LICORNE hall.}
\label{fig:aris_setup}
\end{figure}

The ARIS (Argon Response to Ionization and Scintillation) experiment is a fixed kinematics scattering experiment utilizing a LAr TPC aimed to investigate the response of LAr to nuclear and electronic recoils, with nuclear recoils measured down to $\sim$2~keVee (electron equivalent energy).
The ARIS TPC was exposed to the LICORNE pulsed neutron source at the ALTO facility in Orsay, France \cite{WILSON201431}.  The LICORNE source exploits the \Lipn inverse kinematic reaction, which guarantees a highly collimated and quasi-monoenergetic ($\sim$1.5~MeV) neutron beam, and at the same time, 
monoenergetic gammas from the 478~keV $^7$Li* de-excitation in coincidence with the beam pulse. Neutrons and gammas scattered in the TPC  are detected by an array of eight liquid scintillator detectors (labelled A0 to A7) which constrain the recoil energy in the TPC through the detector angle with respect to the TPC-beam axis. A picture of the setup is shown in figure~\ref{fig:aris_setup}.

In this work, we report on the precise measurement of the LAr scintillation efficiency for nuclear and electronic recoils at null field, and the dependence of the electron-ion recombination effect on the electric field.  

\section{Experimental setup}
\label{sec:setup}

\begin{figure}[t]
\includegraphics[width=0.48\columnwidth]{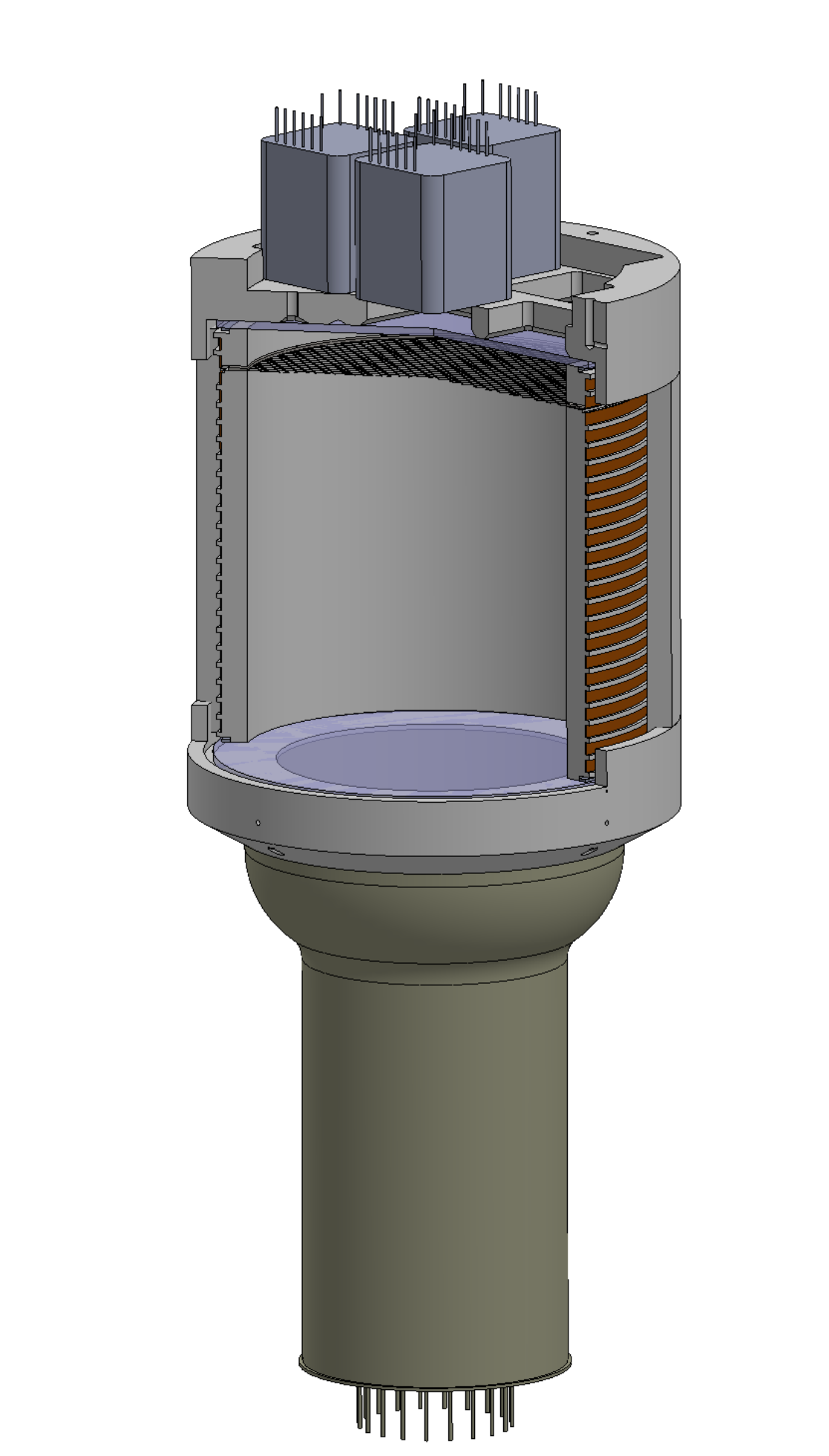}
\includegraphics[width=0.48\columnwidth]{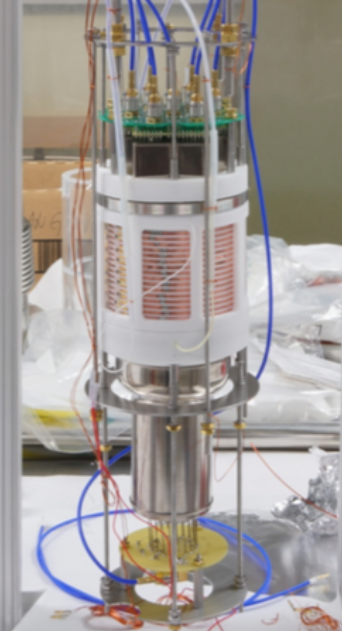}
\caption{Left panel: 3D drawing of the TPC. Right panel: picture of the TPC.}
\label{fig_setup_drawing}
\end{figure}

The ARIS TPC was designed to minimize non-active materials in the direction of the neutron beam to inhibit interaction in passive materials.
The $\sim$0.5 kg  LAr active mass  is housed in a 7.6~cm inner diameter, 1~cm thick Polytetrafluoroethylene (PTFE) sleeve.
The PTFE inner surface includes an embedded Enhanced Specular Reflector (ESR) film for increased light reflection.
The PTFE sleeve supports a set of  2.5~mm thick copper rings connected by resistors in series to maintain a uniform electric field throughout the active argon volume as depicted in figure~\ref{fig_setup_drawing}.  The TPC is held in a double-walled stainless steel dewar.  Evaporated argon is continuously purified with a getter and re-condensed by means of a custom cold head.

The electric field is created by two fused silica windows 
placed at the end-caps of the cylindrical volume.  The fused silica windows are coated with indium tin oxide (ITO) which is a transparent conductor. 
A hexagonal stainless steel grid is placed 
1~cm below the anode to enable the creation of an extraction field for a potential S2 signal.   The anode is held at ground while the voltages on the extraction grid and cathode are tuned to create a uniform drift field across the entire liquid volume.  The electric field uniformity was confirmed with COMSOL \cite{COMSOL:4.4} simulations, with deviations smaller than 1\% for all drift fields.  

For the measurements presented in this paper, the TPC was operated in single phase without a gas pocket, in both field-off/field-on regimes.  This allows for a minimal DAQ acquisition gate, leading to a reduced accidental background.  In a dual-phase TPC, the gate length is dominated by the electron drift time, which depends on the electric field.  By collecting only S1 signals, the gate can be reduced from tens/hundreds of microseconds to $\sim$10~$\mu$s.  The amplitude of the ionization component can be inferred from the comparison of S1 signals with/without the electric field applied, as discussed in section \ref{sec:recombination}.
 

The scintillation photons are wavelength shifted from the ultraviolet to visible range by the tetraphenyl butadine (TPB) compound, which has been evaporated onto all surfaces facing the active volume.  Wavelength shifted photons are observed by one 3-inch R11065 photomultiplier tube (PMT) below the cathode and seven 1-inch R8520 PMTs above the anode.  An optical fiber connected to a LED, powered by a pulse generator, is used to calibrate the single photoelectron response of the PMTs. 

The TPC is mounted with its center 1.00~m away from the LICORNE neutron production target, a hydrogen gas cell which is exposed to a \Li beam that can be accelerated to different energies.
For the measurements presented in this paper the \Li energy was set at 14.63~MeV.  
The gas cell and the beam pipe are separated by a thin tantalum foil where \Li nuclei lose some of their energy. The determination of the \Li energy after the tantalum foil and the parameters of the neutron beam are described in the next section. The \Li beam provides 1.5~ns wide pulses every 400~ns with a current between 20 and 40 nA.  The neutrons reaching the TPC are of the order of 10$^4$~Hz. 

The eight neutron detectors (NDs) surrounding the TPC have active volumes of NE213 liquid scintillator, with a diameter of 20~cm  and an height of 5~cm~\cite{LAURENT1993517}.  The pulse shape of the signal from the liquid scintillator can be used to discriminate between neutrons and $\gamma$s.  The NDs are located at distances from the TPC ranging from 1.3 to 2.5~m, oriented at angles between 25.5 and 133.1 degrees (see table~\ref{tab_NRenergies}). The ND positions were precisely measured before the data taking with a survey method yielding an accuracy of 2-3~mm, depending on the ND.  An inspection after the data taking identified a mismatch between the recorded position of A2 and its position during data taking, which is reflected in a larger systematic uncertainty for measurements using that data point, described in further detail in section~\ref{sec:quenching}.

	\begin{table}
	\centering
	\begin{tabular}{c |  c | c | c}
		\hline
		\hline
					&	 Scattering	&	Mean NR			&	Mean ER \\
					&	Angle [deg]	&	Energy [keV]		&	Energy [keV]\\
       		\hline 
			A0		&	25.5			&	7.1	                 & 42.0                  \\
			A1		&	35.8			&	13.7		         & 75.9                  \\
			A2		&	41.2			&	17.8		         & 85.8                  \\
			A3		&	45.7			&	21.7		         & 110.3                  \\
			A4		&	64.2			&	40.5		         & 174.5                 \\
			A5		&	85.5			&	65.4		         & 232.0 		   \\
			A6		&	113.2		&	98.1		        & 282.7			\\
			A7		&	133.1		&	117.8		& 304.9              \\
		\hline
		\hline
	\end{tabular}
	\caption{Scattering angles, NR mean energies for neutrons from the \Lipn reaction, and ER mean energies from Compton scattered $\gamma$s emitted by \Lim de-excitation, are shown. The scattering angle is defined with respect to the center of the NDs active surface while the mean energies are determined with Monte Carlo simulations.}
	\label{tab_NRenergies}
	\end{table}
 
Data taking occurred during a 12 day period in October 2016 with various electric fields in the TPC, ranging from 0 to 500~V/cm. Data were taken in two modes: double coincidence mode  between the beam pulse and a TPC trigger, and a triple coincidence mode which included also coincidence with at least one of the NDs. The TPC trigger condition requires at least  two PMTs to fire within 100 ns and a measurement of the TPC trigger efficiency will be described in section~\ref{sec_calibration}.  The triple coincidence data set provides nuclear and electronic recoils of defined energies. 
The double coincidence data, which provides continuous spectra, are used for an investigation of the LAr scintillation time response, which will be presented in a future publication. 

When a trigger occurs signals from the  TPC PMTs and from A0--A7 are digitized by two CAEN V1720 boards at a 250~MHz frequency. The time of the beam pulses is also digitized at a 250~MHz frequency by a CAEN V1731 board.  The board timestamps are synchronized by an external clock to allow for time-of-flight measurements. 

For each coincidence, the TPC PMT waveforms, the ND waveforms, and the signal from the beam pulse are recorded.  The acquisition window was 10~$\mu$s for the TPC PMTs and 7~$\mu$s for each ND. The signals are analyzed by a reconstruction software based on the \textit{art} framework \cite{Green:2012gv} to extract observables from the recorded waveforms.  First, fluctuations and drift of the baseline are tracked and subtracted from the raw signal waveforms.  Next, waveforms from each PMT in the TPC are corrected for their single photoelectron response and summed together.  A pulse finder algorithm is applied to each summed waveform to identify the magnitude and start time of TPC and ND pulses.  Finally, the reconstructed waveform and pulse information are used to extract the \SOne amplitude, pulse shape discrimination parameters for both the TPC and NDs, and time-of-flight (TOF) parameters.  

A Geant4-based Monte Carlo simulation of the experimental setup has been developed which includes the materials, size, and relative placement of the TPC, PMTs, dewar, and A0--A7 detectors as described above.  The beam kinematics is also included as described in section~\ref{sec:BeamKinematics}. This simulation provides a spectrum of nuclear and electronic recoil energies from coincidences between the TPC and A0--A7 detectors, with mean values listed in table~\ref{tab_NRenergies}.


\section{Neutron beam kinematics}
\label{sec:BeamKinematics}

The LICORNE neutron beam exploits the inverse kinematic reaction resulting from accelerated \Li incident on a gaseous hydrogen target.  The kinematics of the neutrons emitted from the \Lipn reaction highly depends on the energy of the \Li at the reaction site.  The \Li beam is initially accelerated to 14.63~MeV, and a fraction of its energy is lost as it crosses the tantalum foil containing the hydrogen target.  The exact thickness of the foil, and therefore the final \Li energy, is not well known.  A dedicated measurement was performed to determine the \Li energy at the reaction site, and therefore the kinematic profile of the neutron beam.  One ND was placed at a distance of 3~m from the source at angles varying between 0 and 15$^\circ$.  The relative neutron beam intensity with respect to the intensity at 0$^\circ$ was measured at each angle.  The resulting profile was compared with the results of a GEANT4 simulation developed to predict neutron spectra as a function of \Li energy \cite{Lebois:2014il}, which assumes different thicknesses of the tantalum foil. 

\begin{figure}
\includegraphics[width=\columnwidth]{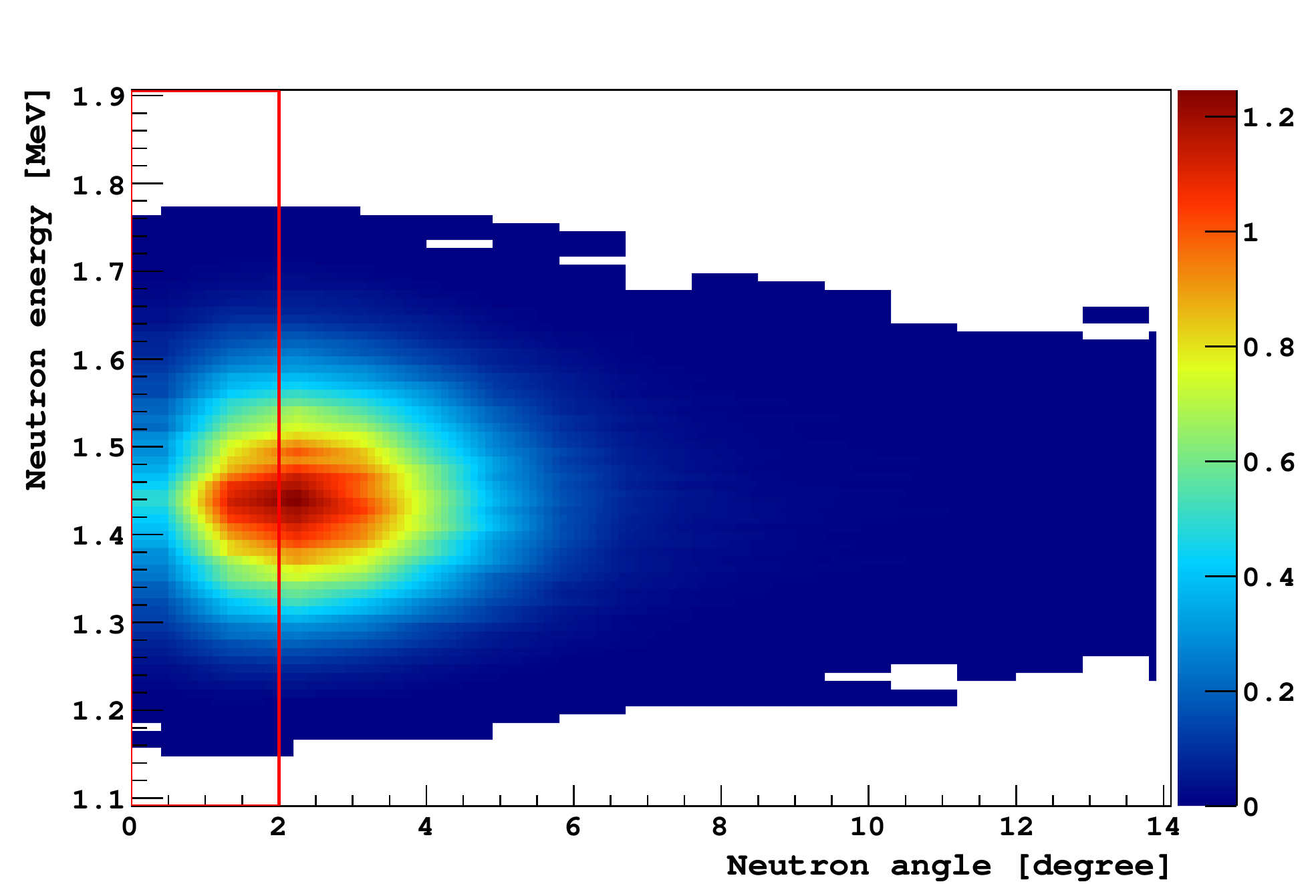}
\caption{Neutron kinematic profile for a \Li energy of  13.13~MeV incident on the hydrogen target, determined from Monte Carlo simulations. The red box defines the geometrical acceptance of the TPC. } 
\label{fig_calib_kine_bestfit}
\end{figure}

The best fit of the  beam profile was obtained for a foil thickness of $2.06 \pm 0.08\,\mathrm{\mu m}$, corresponding to a mean \Li energy in the hydrogen target of $13.13 ^{+0.02}_{-0.01}\,\mathrm{MeV}$. 
The corresponding kinematic profile of the neutron beam, in neutron energy vs. angle with respect to the \Li beam axis, is shown in figure~\ref{fig_calib_kine_bestfit}. 
The TPC, located 1 m from the neutron source, is exposed to $<$2$^\circ$ of the neutron cone. 
The mean neutron energy in this region is 1.45~MeV with an RMS of 85~keV.

In addition to neutrons, LICORNE isotropically emits a source of 478 keV $\gamma$s in the center of mass frame from either  the \Be decay or by the de-excitation of \Lim,  produced when \Li crosses the tantalum foil. 
The decay of \Be, with a half-life of $\sim$53 days, constitutes a source of constant accidental background within the beam pulse.
The \Lim $\gamma$s are emitted in coincidence with the beam pulse, and when detected in coincidence between the beam pulse, TPC, and a ND, provide an excellent source of single Compton electrons for investigating the LAr response to ERs. 
The \Lim $\gamma$s are subjected to a relativistic boost due to the motion of the \Lim nuclei, which increases their energy up to 6\% for a \Li energy of 14.63~MeV.  Since the \Lim energy at which  $\gamma$s are emitted can vary because of the energy loss in the source materials, a mean boost of  3\% and,  conservatively, a $\sigma$ of 3\% are assumed, resulting in  a $\gamma$ energy of 492$\pm$15~keV.


\section{Detector calibration}
\label{sec_calibration}

The TPC response to scintillation light was calibrated throughout the data taking period with  \am\ and \ba $\gamma$-sources placed on the outside surface of the dewar.  Dominant $\gamma$-lines able to cross the dewar walls and reach the active LAr mass are  59.5~keV from  \am  and  81.0~keV to 383.8~keV from \ba. 

\begin{figure}[!t]
\includegraphics[width=\columnwidth]{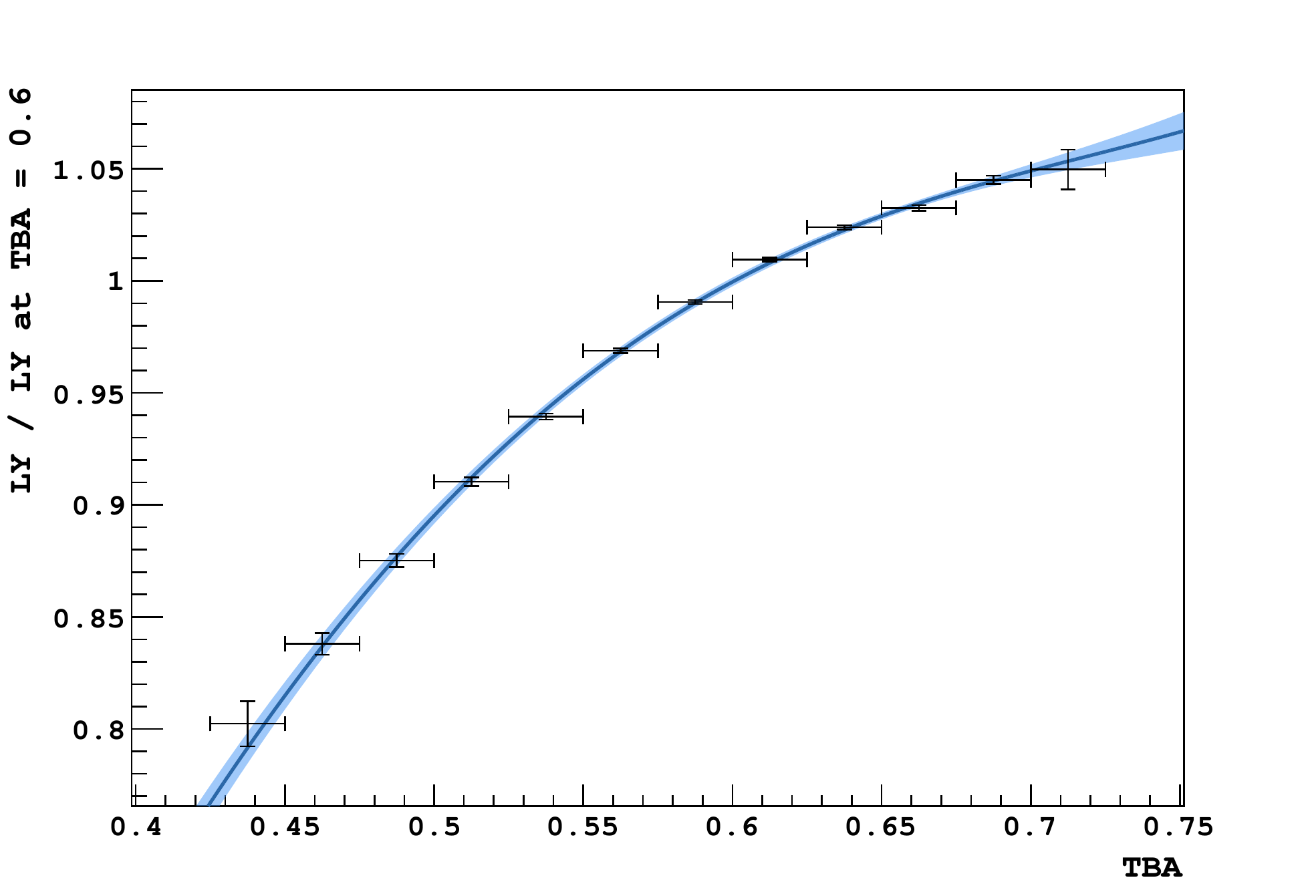}
\caption{Relative light collection efficiency as function of the TBA. Larger values of TBA correspond to recoils closer to the bottom of the TPC.}
\label{fig:TBA}
\end{figure}

\begin{figure}[!htpb]
\includegraphics[width=\columnwidth]{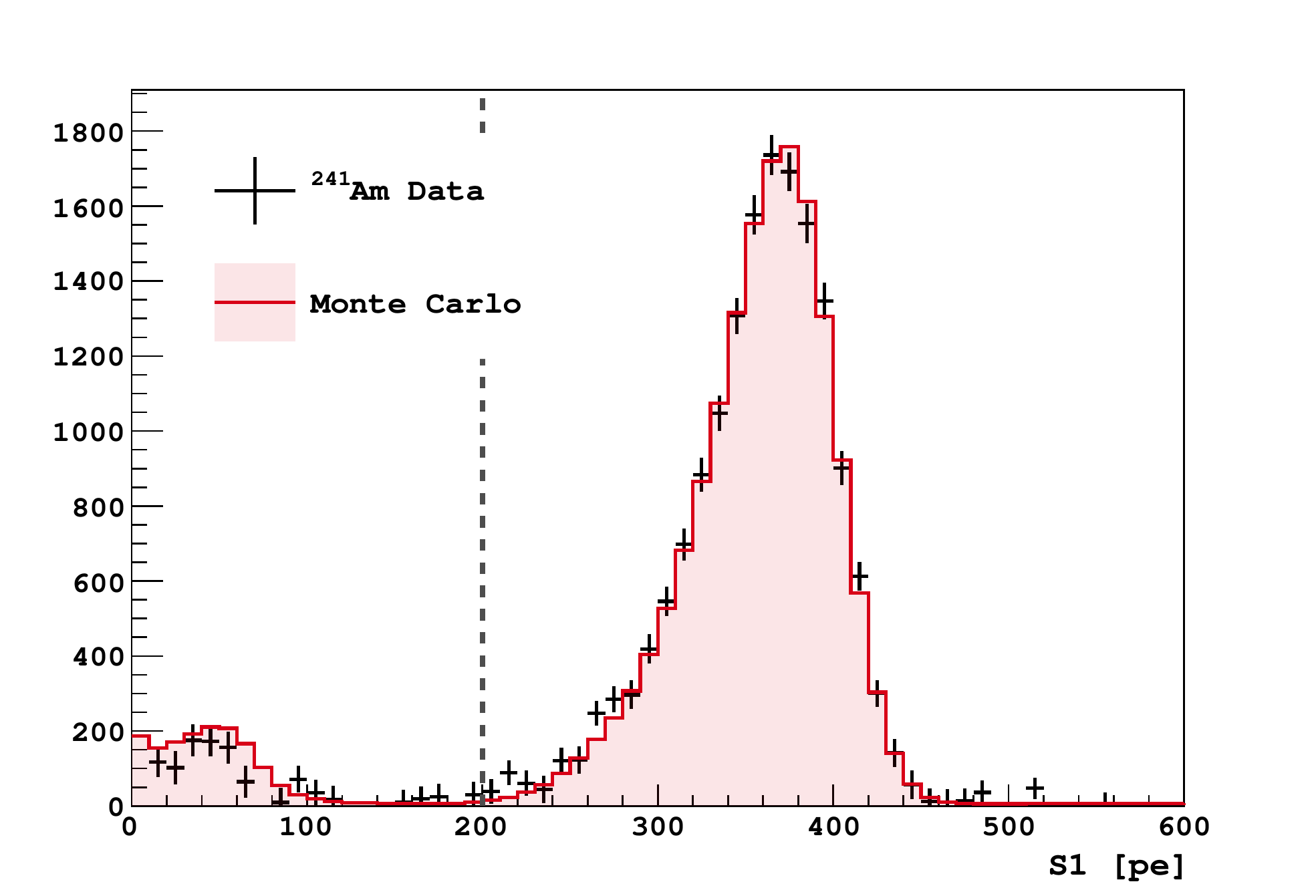}
\includegraphics[width=\columnwidth]{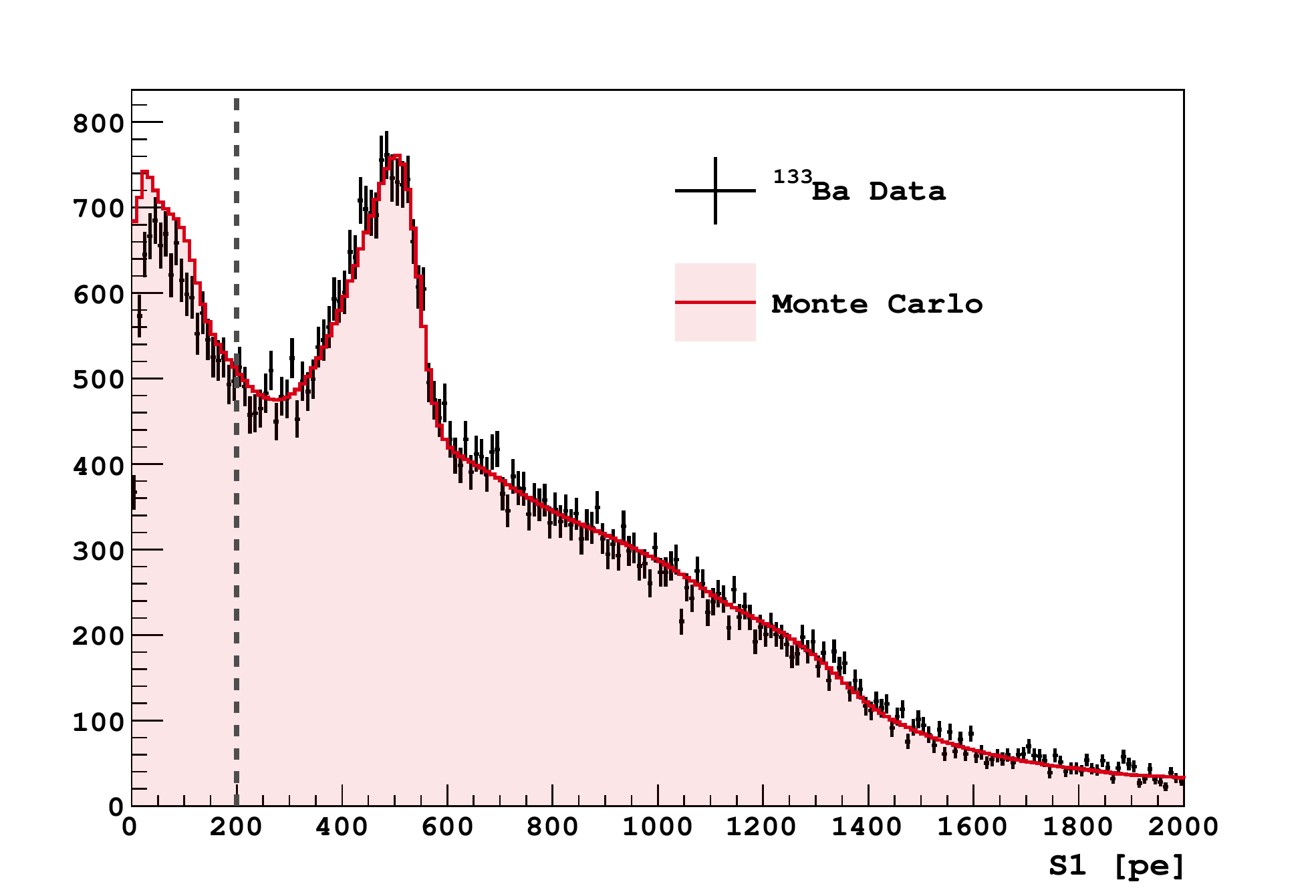}
\caption{\am (top) and \ba (bottom) spectra from source calibration of the TPC light-yield at 0~V/cm with an overlay of the best fit spectra. The $\chi^2/n.d.f$ is 20.7/29 for \am and 190.7/189 for \ba. The vertical dashed line represents the low threshold for the fit interval.}
\label{fig_calib_ly}
\end{figure}

The energy deposits of the $\gamma$s in the LAr active target are simulated with a Geant4-based Monte Carlo. Its output is converted to the S1 observable by convolving energy deposits in the LAr with a response function. The response function is generated with a toy Monte Carlo approach, taking into account  the light yield (LY),  Poisson fluctuations in the photon statistics, the non-uniformity of  the light collection along the vertical axis, and the PMT response. The last is measured by fitting 
the single photoelectron distribution obtained with a pulsed LED light fed to the TPC through an optical fiber.
The single photoelectron distribution was monitored throughout the data taking period.
The topological uniformity in light collection is measured by looking at the TBA (top/bottom asymmetry)  observable, defined as the ratio between the light collected by the bottom PMT with respect to the total.  The light collection is expected to be larger at the bottom because the 3-inch PMT provides a larger  optical coverage and   quantum efficiency with respect to the 1-inch PMT array at the top. 
The $\gamma$s from the \am source are used to evaluate the dependence of the light collection  on the TBA.
In figure~\ref{fig:TBA}, the mean S1 of the \am $\gamma$ peaks observed in different subranges of TBA is shown with respect to S1 for TBA=0.6. TBA=0.6 corresponds to the mean value of the TBA for events induced by a source placed at the center of the TPC.

\begin{figure}[!htbp]
\includegraphics[width=\columnwidth]{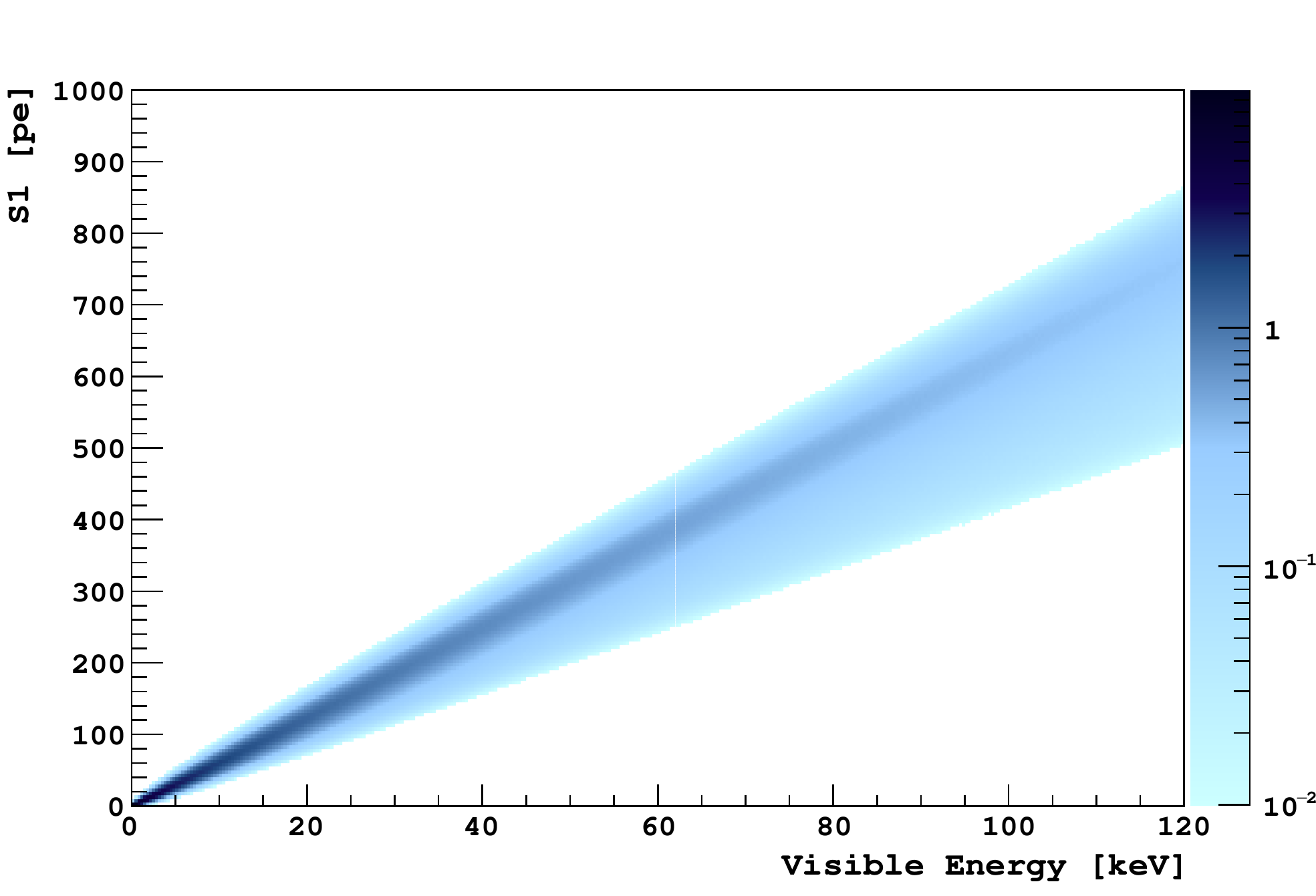}
\caption{The TPC optical response as a function of visible energy defined from the toy Monte Carlo approach described in the text.}
\label{fig_response_map}
\end{figure}

The \am and \ba data are then fit with the simulated distributions with the LY as the only free parameter.  The \am and \ba spectra and best fit Monte Carlo distributions are shown in figure~\ref{fig_calib_ly}.  A $\chi^2/ndf\sim1$ is achieved for both sources, showing that  data and Monte Carlo are in excellent agreement. 
A slow 1.8\% decrease in the LY over the period of data taking, likely due to variations in the LAr purity,  was observed with daily calibrations with the $\gamma$ sources, resulting in the dominant systematic error on the LY measurement.
The best-fit is obtained for LY~=~6.35$\pm$0.05~pe/keV with the uncertainty including the systematic error on the LY stability.
The response map obtained for the average LY is shown in figure~\ref{fig_response_map}.

The trigger efficiency is derived with a dedicated measurement with a \Na source placed on the external wall of the dewar.  \Na emits a positron, resulting in two back-to-back 0.511~MeV $\gamma$s, simultaneous with an isotropic 1.27~MeV $\gamma$. Two BaF$_2$ detectors were  placed at a distance of $\sim2$~cm from the source: one on the TPC--\Na source axis in order to detect one of the two 0.511~MeV $\gamma$s from the positron annihilation, and the second  rotated by about 45 degrees with respect to the same axis to detect the isotropic 1.27~MeV $\gamma$. 
This trigger configuration provides selection of events where one 0.511~MeV $\gamma$ is directed toward the TPC center 
when both BaF$_2$ detectors are triggered.  In this case, the TPC event and  the TPC trigger status (two or more PMTs fire within 100 ns) are recorded along with the light  collected in the two BaF$_2$ detectors. 
Offline  cuts on the BaF$_2$ signals optimize the selection of  0.511~MeV $\gamma$-rays directed toward the TPC center, inducing a  Compton electron spectrum ensuring a trigger efficiency scan over the entire energy range of interest.  A dependence of the trigger efficiency on the TBA is expected due to the asymmetry in the photosensor setup on the top and bottom of the TPC and the trigger condition.

\begin{figure}[t]
\includegraphics[width=\columnwidth]{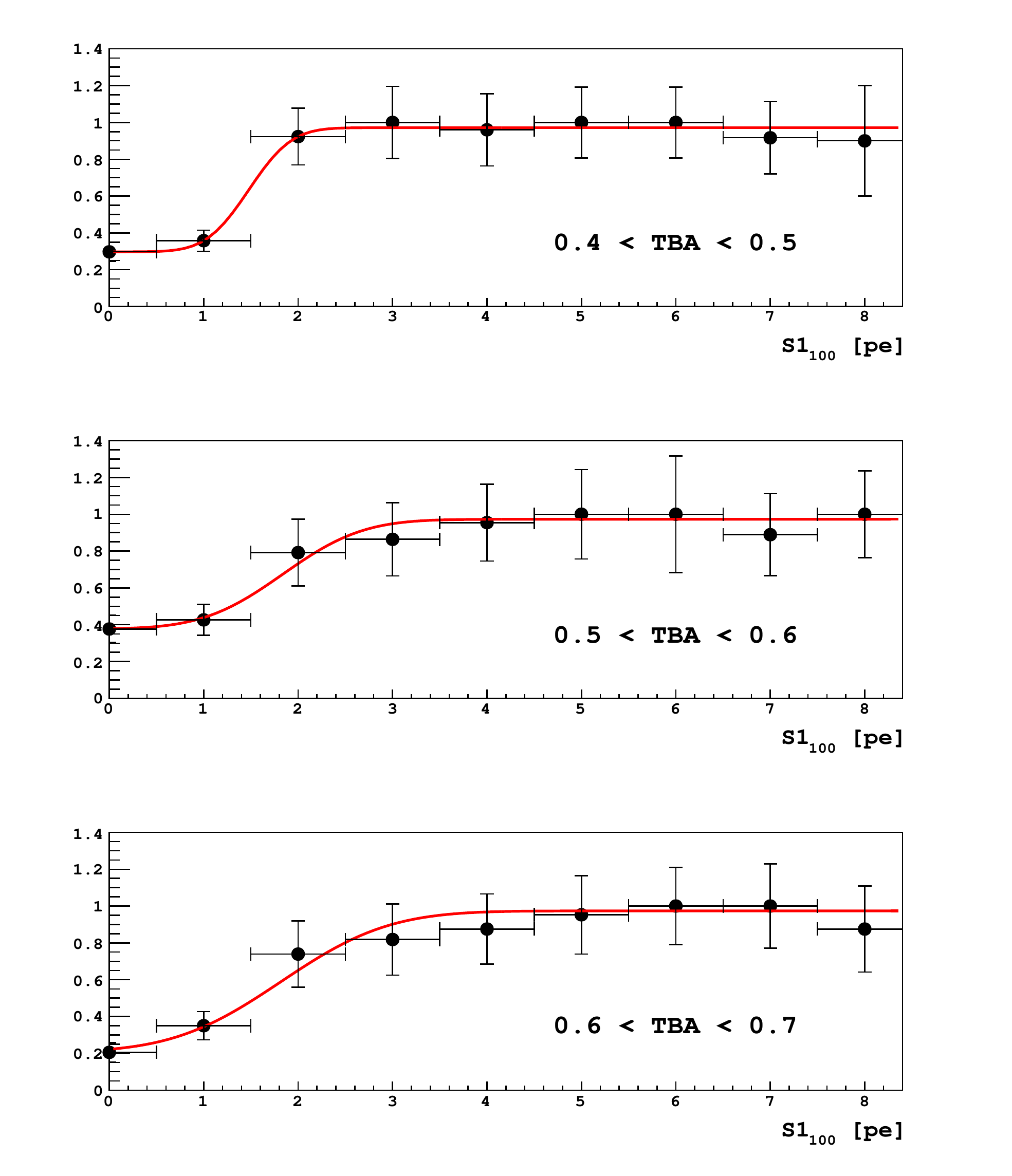}
\caption{ The trigger efficiency, as a function of S1$_{100}$, measured with the $^{22}$Na source for three regions of TBA.   The  plateau at  high energies does not reach unity due to the inhibition time of 10 ms introduced after each trigger. Dark noise prevents the efficiency from reaching zero at  very low values of S1$_{100}$. }
\label{fig_trigg_eff}
\end{figure}

\begin{figure}[t]
\includegraphics[width=\columnwidth]{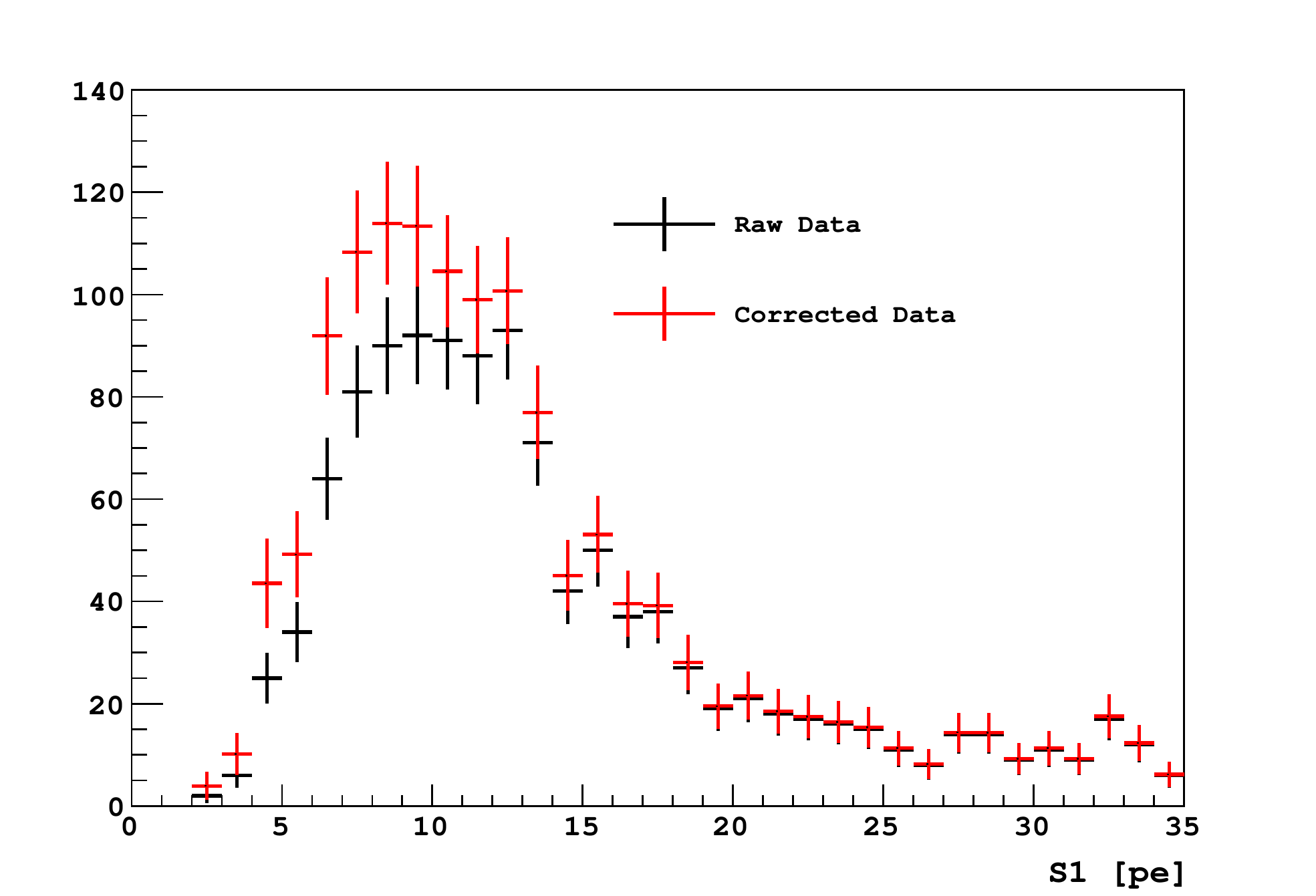}
\caption{The effect of the trigger efficiency correction on the NR energy spectrum for events selected by coincidence with A0, resulting in a mean NR energy of 7.1~keV$_{nr}$.  The impact of the trigger efficiency is maximal for this data point. }
\label{fig_A0_trigger}
\end{figure}

The trigger efficiency is measured as a function of the signal in the first 100 ns (S1$_{100}$), the same gate as the one used for the trigger. Monte Carlo simulation have demonstrated that ER and NR have the same trigger efficiency with respect to this variable.
The fraction of reconstructed events with a positive trigger status as a function of S1$_{100}$ is evaluated for  three regions of TBA, approaching 1 in the entire detector for S1$_{100}$ $>$6, as shown in figure \ref{fig_trigg_eff}. Beam data are corrected on an event-by-event basis by evaluating the corresponding S1$_{100}$ value.
Figure \ref{fig_A0_trigger} shows the comparison of corrected and un-corrected NR spectra selected by coincidence with the A0 detector (7.1~keV$_{nr}$), where the impact of the trigger efficiency is maximal.

The TPC saturation has been investigated with the \Na source by comparing S1 with S1$_{late}$, the integral of the signal starting after the first 90~ns. This integral range is not affected by saturation since it is dominated by the slow component of the scintillation emission with a characteristic time of $\sim$1.6~$\mu$s. A deviation from the linearity between S1 and S1$_{late}$ is observed from S1=4000~pe, corresponding to more than 600~keV$_{ee}$. A similar study has been performed on the spectrum of NRs selected in the double coincidence trigger mode.  The prompt scintillation component in NRs is larger than for ERs, so the effect of saturation is expected at lower S1. Up to 400~pe, corresponding to the maximum energy of NRs induced by 1.45~MeV neutrons, no deviations from linearity were observed between S1 and S1$_{late}$.

\begin{figure}[t]
\includegraphics[width=\columnwidth]{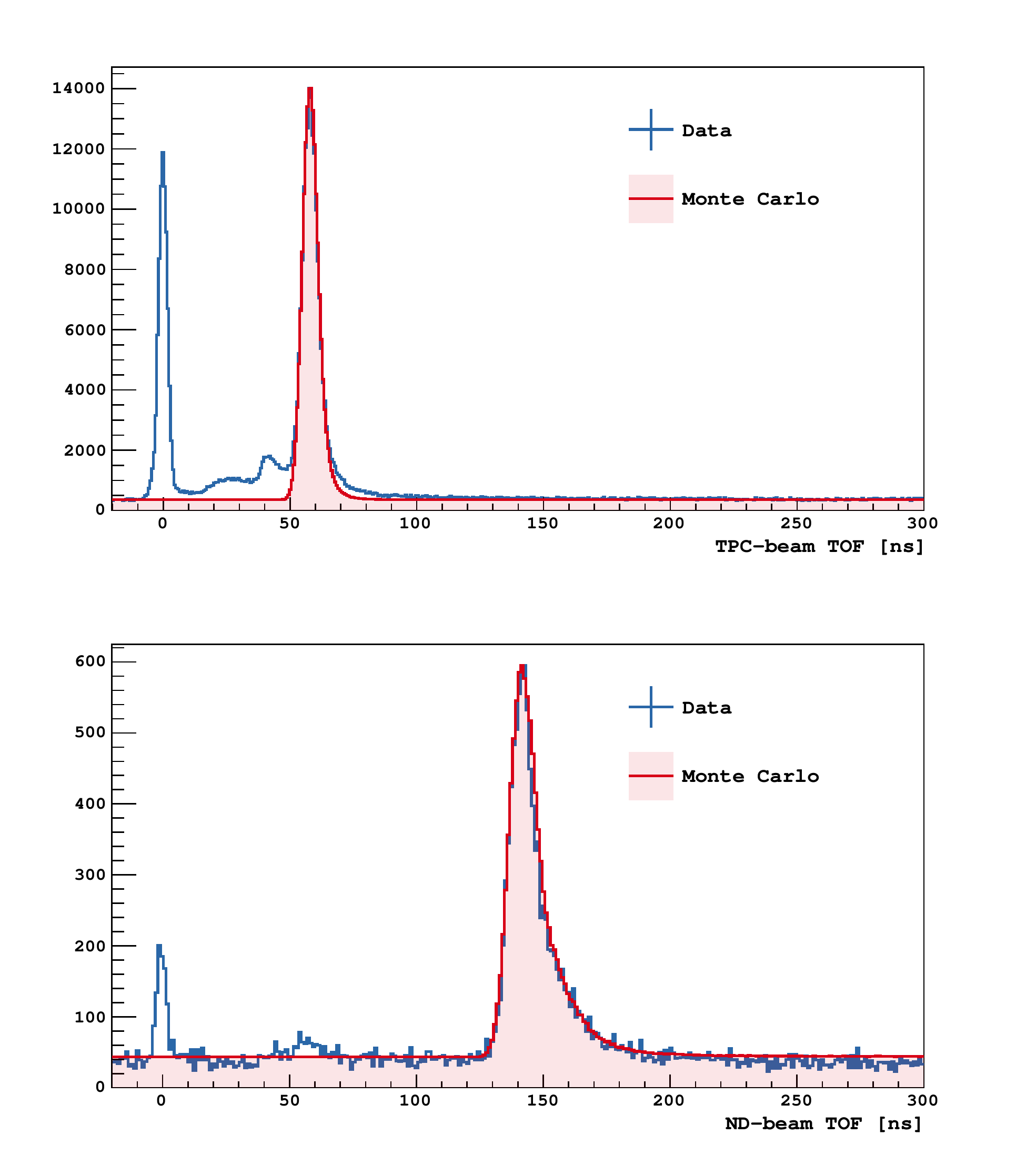}
\caption{Data and Monte Carlo comparisons of the \TPCtof (top) and  \NDtof (bottom) distributions for the A3 detector. The peak at 0~ns corresponds to the \Lim $\gamma$ that are not simulated in the MC. 
}
\label{fig_tof_mc-data}
\end{figure}

A clock misalignment was occasionally observed on a run-by-run basis between the CAEN boards. To synchronize them,  a time calibration for each run was performed using the \Lim $\gamma$ signal from the triple coincidence trigger data. 
The TOF between the beam pulse and the TPC (\TPCtof) and the beam pulse and the NDs  (\NDtof) is shown in figure \ref{fig_tof_mc-data}, compared with  Monte Carlo simulations of neutrons for the A3 ND.  The time resolutions of the two TOFs are measured with  \Lim $\gamma$s to be 1.8 ns for the \TPCtof, and in the 2--3~ns range for the eight \NDtof and are included in the simulation. 
The excellent agreement in both TOF distributions is an indirect confirmation of the neutron kinematic profile assumed in the Monte Carlo.

\section{Selection criteria}
\label{sec:selection}


In order to understand the different populations  of events in the the data sample, four observables are used:  the two previously defined TOF variables (\TPCtof and  \NDtof), and the TPC  (\FNine) and  ND (\NDpsd) pulse shape variables.  \FNine   is defined as the fraction  of the first 90~ns of the light pulse in the TPC, while  \NDpsd corresponds to the fraction of photoelectrons detected  after 40~ns up to the end of the acquisition gate (7~$\mu$s) in the NDs.

\begin{figure}[t]
	\includegraphics[width=\columnwidth]{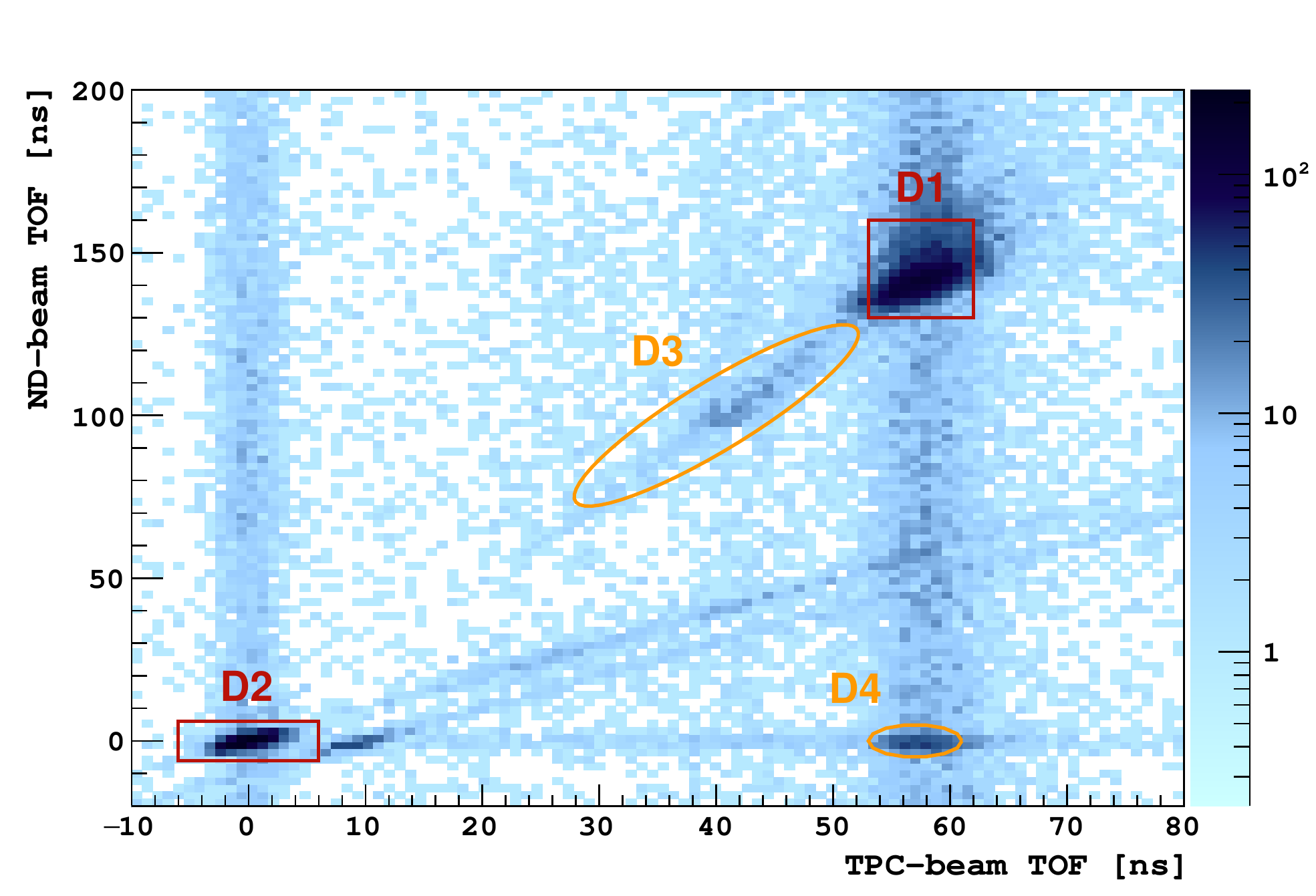}
	\includegraphics[width=\columnwidth]{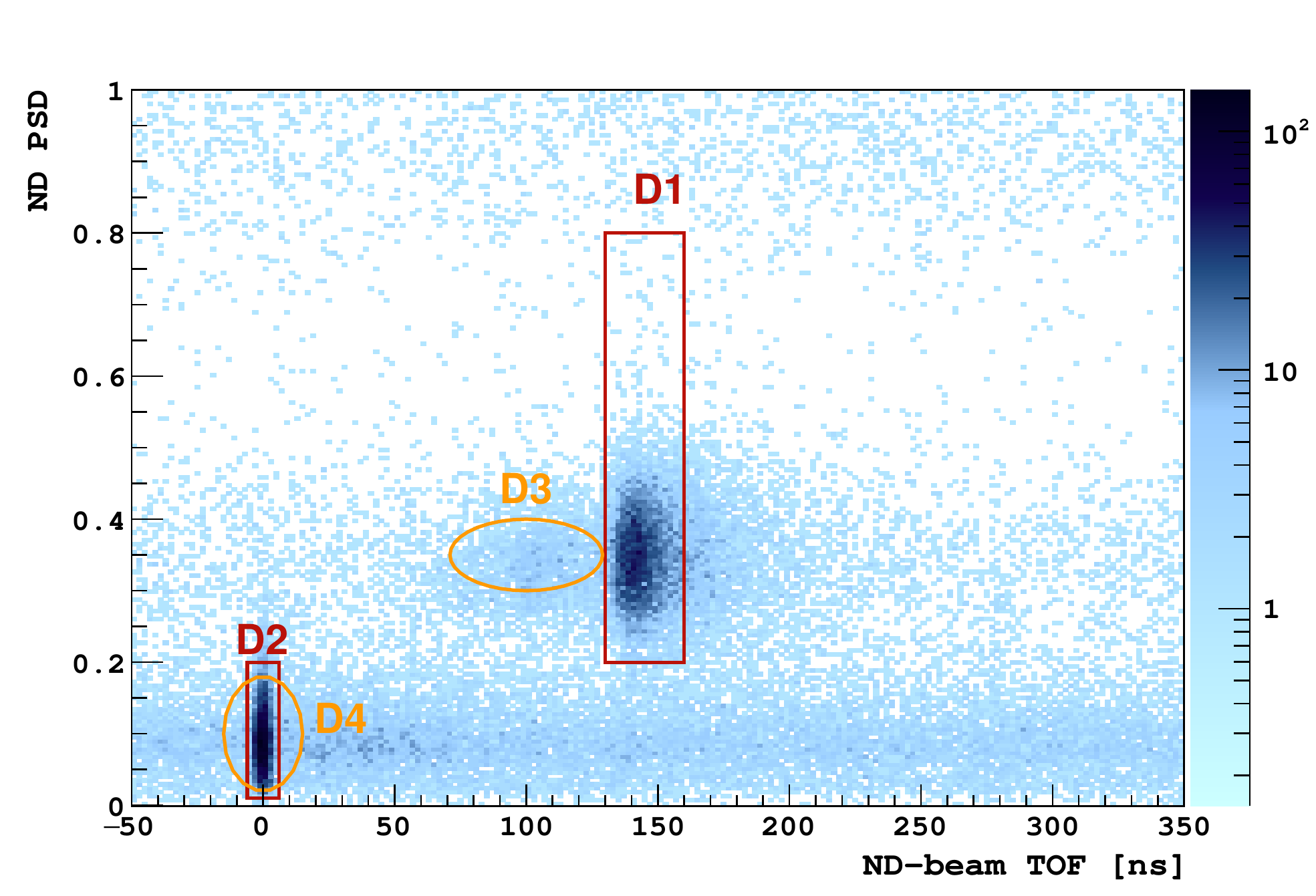}
	\includegraphics[width=\columnwidth]{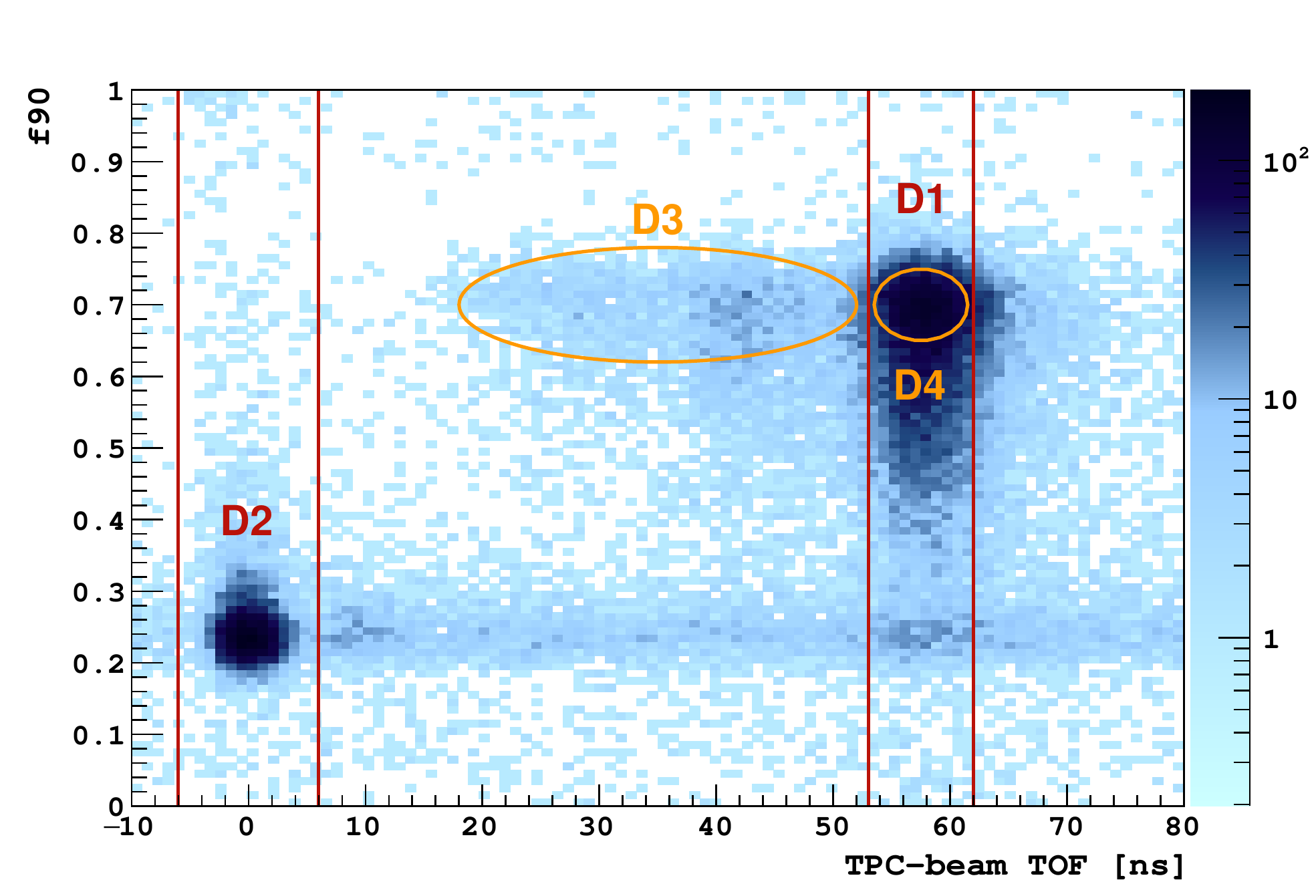}
	\caption{ \NDtof vs. \TPCtof (top),  \NDpsd vs. \NDtof (center), and \FNine vs. \TPCtof (bottom) for triple coincidences with the A3 detector.  The numbered populations are described within the text. The red lines corresponds to the selection cuts for NRs (D1) and ERs (D2). Yellow lines highlight two classes of NR (D3) and ER backgrounds. }
	\label{fig_selection}
\end{figure}

Figure~\ref{fig_selection} shows different combinations of   observables  for triple coincidence events in the A3 ND,   highlighting four different classes of events as well as the selection cuts for NR and ER signals.  The four classes of events described in the subsequent paragraphs are labeled in the figure.

{\bf D1} Large \FNine indicates that these events are mostly NRs, and the values of \mbox{\TPCtof$\sim60$~ns} and \mbox{\NDtof$\sim150$~ns} are in agreement with the  expected TOF from $\sim1.5$~MeV neutrons  traveling 1~m to the TPC    and 2.5~m to the ND.  The    \NDpsd  variable  confirms that these events are neutrons,  with a mean value of  $\sim0.35$ corresponding to the expectation for proton recoil in scintillator from a neutron interaction. 

{\bf D2} The  small \FNine and \NDpsd in combination with TOF values at the few nanosecond scales provide clear indications that these events are beam-generated $\gamma$s interacting in both the TPC and ND.  

{\bf D3} \FNine classifies   these events as  neutrons, but the  two TOFs are shorter  than for the expected signal from   $\sim$1.5~MeV neutrons. These high energy neutrons are identified as byproducts of fusion-evaporation reactions between the different target materials and the accelerated \Li. 

{\bf D4} The short \NDtof, compatible with   $\gamma$s in the ND, and the long \TPCtof, compatible with neutrons in the TPC,  identify these events as accidental coincidences between a neutron and $\gamma$ correlated with the beam pulse.  

Events in D1 and D2 categories are selected by combining cuts on  \TPCtof, \NDtof, and \NDpsd variables. The selection has been optimized independentely for each ND. As an example, the cuts used for A3 are shown by the red boxes in figure~\ref{fig_selection}. 
The \FNine cut is excluded by the data selection to avoid possible biases in the TPC energy spectra due to the correlation between S1 and \FNine. 

\begin{figure}
\includegraphics[width=\columnwidth]{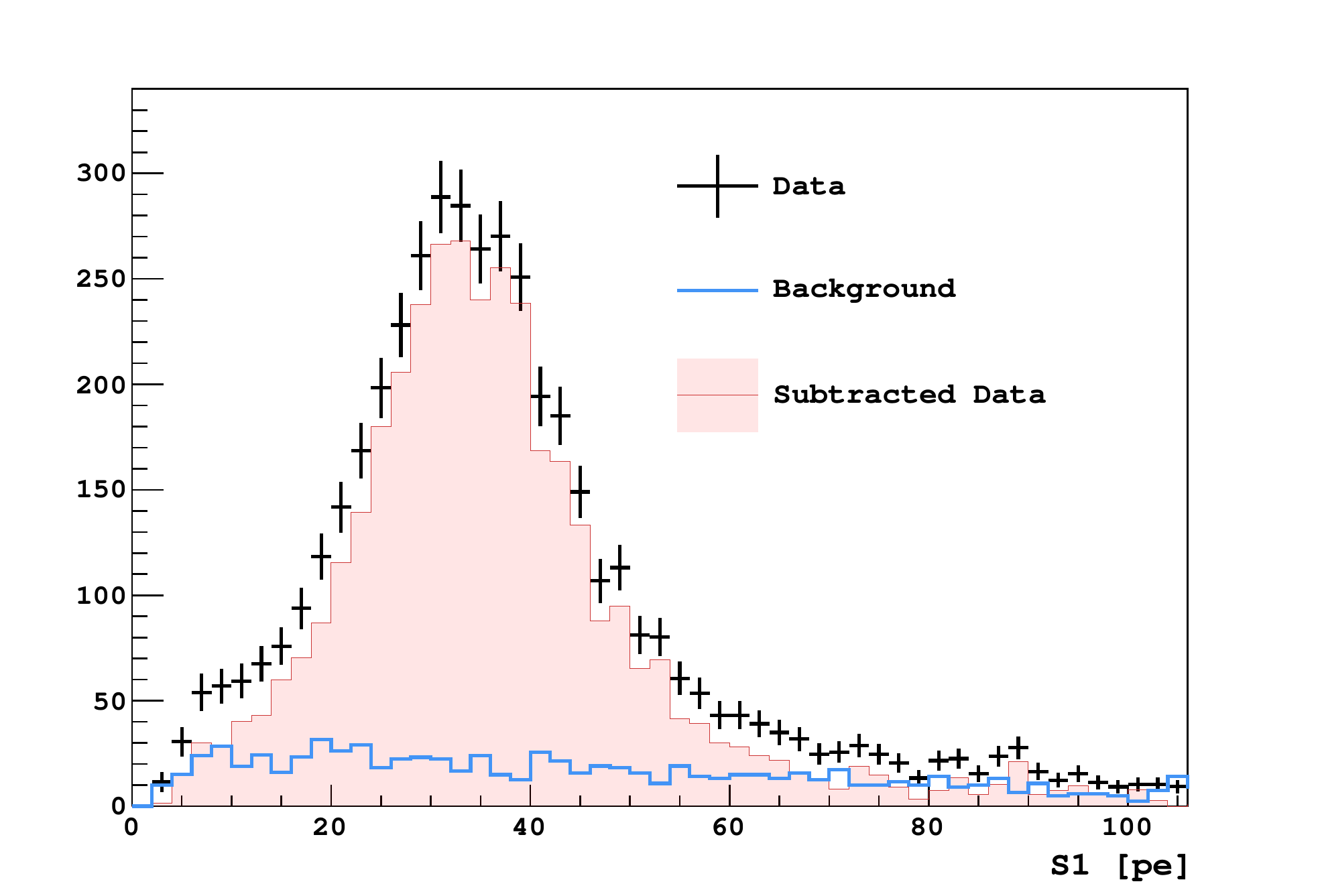}
\includegraphics[width=\columnwidth]{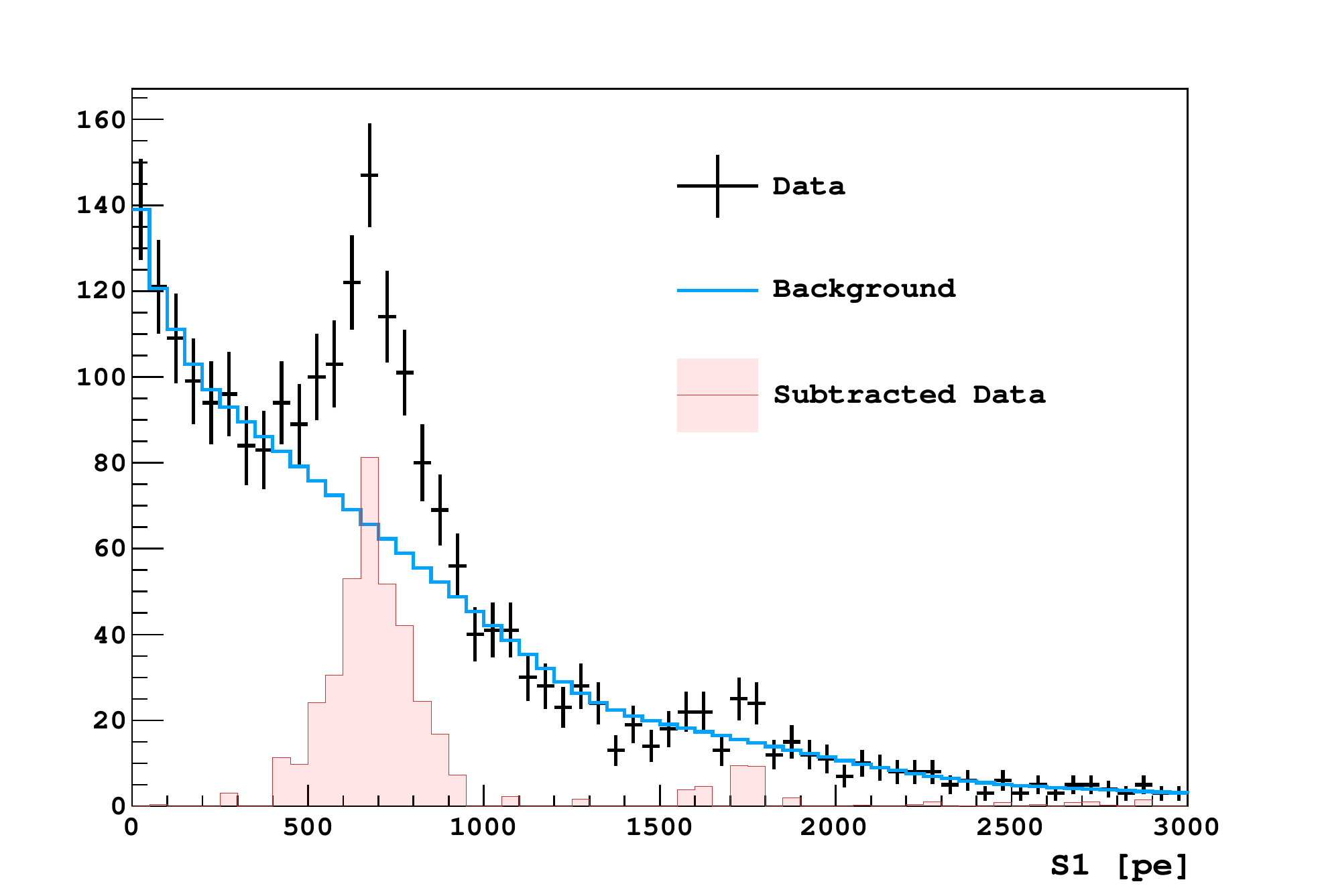}
\caption{Top: NR spectrum in A3 (21.5 keV recoil energy). Bottom: ER spectrum for  Compton electrons in  A3. In black is the spectrum after all the selection cuts. The background is represented by the blue histogram.}
\label{fig_spbkg}
\end{figure}

The most significant background to the neutron signal population is from accidental coincidences between a neutron in the TPC and an ambient $\gamma$ in a ND.  A probability density function of the S1 spectrum of this background is produced by selecting events with the same \TPCtof as is used for the D1 signal events, accepting all events in the \NDtof variable that are not coincident with the D1 or D2 region windows.  The background spectrum is normalized to a high energy region of the signal S1 spectrum for NR events, by requiring S1$>500$~pe, and subtracted, as shown in the top plot of figure~\ref{fig_spbkg} for the ND A3.


In the case of ERs,  $\gamma$s scattering multiple times in the TPC materials is the dominant background, making accidental background subtraction irrelevant. The overall background is   estimated using the \textit{TSpectrum Background} algorithm from \textit{ROOT}~\cite{Brun:1997pa}, as shown in the bottom plot of figure~\ref{fig_spbkg} for  A3.

\section{Response to electron recoils at null field}
\label{sec:linearity}

Existing measurements  \cite{Lippincott:2010jb} suggest that the LAr response at null field to ERs is linear and hence not subjected to energy dependent quenching effects, unlike what has been observed in liquid xenon~\cite{Aprile:2006kx}.  Previous measurements of the linearity of the ER response in LAr have relied on multiple-scatter sources, such as the multi-step decay of \mKr and  $\gamma$ sources in the Compton-scattering dominated regime.  Direct measurements from single electronic recoils have not yet confirmed the linearity.

The eight single ER energies from Compton scattering of the mono-energetic $\gamma$ emitted by the \Lim de-excitation, tagged with the eight NDs, are ideal candles for this test.  
Data from the TPC is background subtracted as described in section~\ref{sec:selection}, and the resulting peak is fit with a Gaussian function. The LY  for Compton scatters tagged by NDs is evaluated  as a function of the  Compton electron energy determined with Monte Carlo simulation. 
The LY of each ND tagged dataset is found relative to the mean value of the set of eight measurements, and the relative LYs are fit with a first degree polynomial resulting in a maximum deviation from unity of 5\%   in the [41.5, 300]~keV range.
The value of this deviation includes the statistical error from the fit. 

\begin{figure}[t]
\includegraphics[width=\columnwidth]{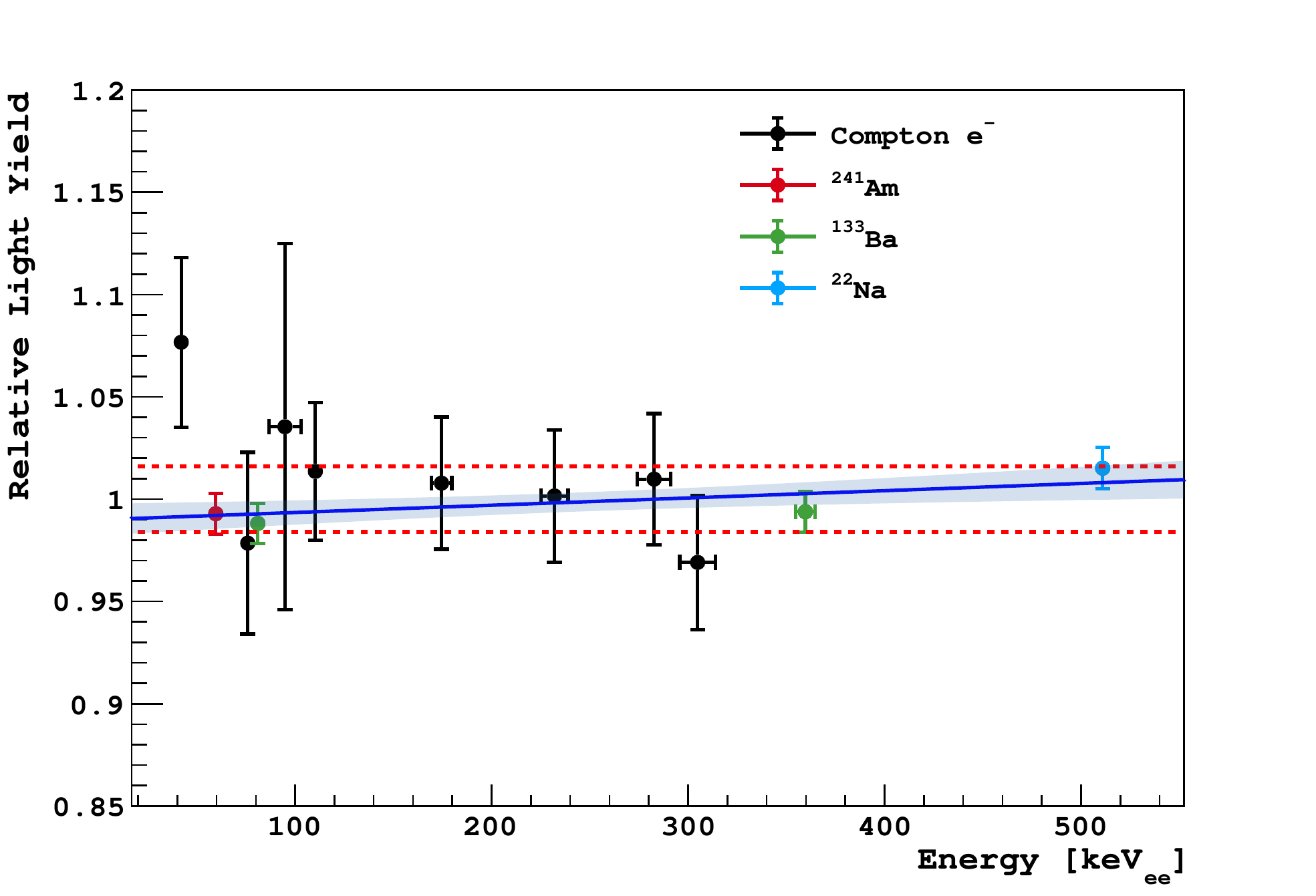}
\caption{
The relative  LY,  with respect to the mean,  as a function of the Compton electron energy from \Lim de-excitation, and from \am (59.5~keV), \ba (81 and 356~keV), and \Na (511~keV) $\gamma$ sources. The vertical error bars represent the statistical errors on LY while the horizantal error bars represent the uncertainties on the energy. All the data points are fit with a first degree polynomial (blues line)  to test for deviations from unity. The dashed red lines correspond to $\pm$1.6\% band and contain the fitted polynomial, including 1$\sigma$ error (blue band), in the [41.5, 511]~keV range.    }
\label{fig_calib_ly_compton}
\end{figure}

The LY values independently extracted from the full absorption $\gamma$ peaks, shown in figure~\ref{fig_calib_ly_compton},  from  \am (59.5~keV), \ba (81 and 356~keV), and \Na (511~keV), are fully compatible with the average LY derived from single Compton electrons. This is expected for the full absorption peaks of 59.5 and 81~keV $\gamma$s as they are dominated by the photoelectric effect. The 356 and 511~keV $\gamma$ interactions,  instead, are dominated by Compton scattering,  producing multiple lower energy electrons.
If they were subjected to a quenching stronger at low energies (as it is the case, for example, for organic liquid scintillator), the light yield derived from multiple scatter $\gamma$s should differ from the one derived from single scatter events.
 
Fitting simultaneously the \am, \ba, \Na, and Compton electrons, the LY in the  [41.5, 511]~keV range is constant within 1.6\%, as shown in figure~\ref{fig_calib_ly_compton}. This result confirms the linearity of the LAr scintillation response at null field also observed by Lippincott \textit{et al.}~\cite{Lippincott:2010jb}  at 3\%, using multiple scatter sources in the [41.5, 662]~keV range.  This result suggests    that, at null field,  ERs are not subjected to non-linear quenching effects and that   calibrations of LAr detectors can be  performed either with single or multiple scatter ER sources without introducing any bias.

\section{Response to nuclear recoils at null field}
\label{sec:quenching}

The scintillation efficiency for nuclear recoils, $\mathcal{L}_{eff}$, is defined in this work with respect to the response of LAr to the 59.5~keV $\gamma$ from \am at null field. 
The comparison with other $\mathcal{L}_{eff}$ measurements~\cite{Gastler:2010sc,Cao:2015ks,Creus:2015fqa} using different reference sources (e.g. $^{57}$Co and \mKr) is guaranteed by the linearity of the ER response demonstrated in the previous section.  

\begin{centering}
\begin{figure*}[!ht]
\begin{minipage}{\textwidth}
	\includegraphics[width=2.80in]{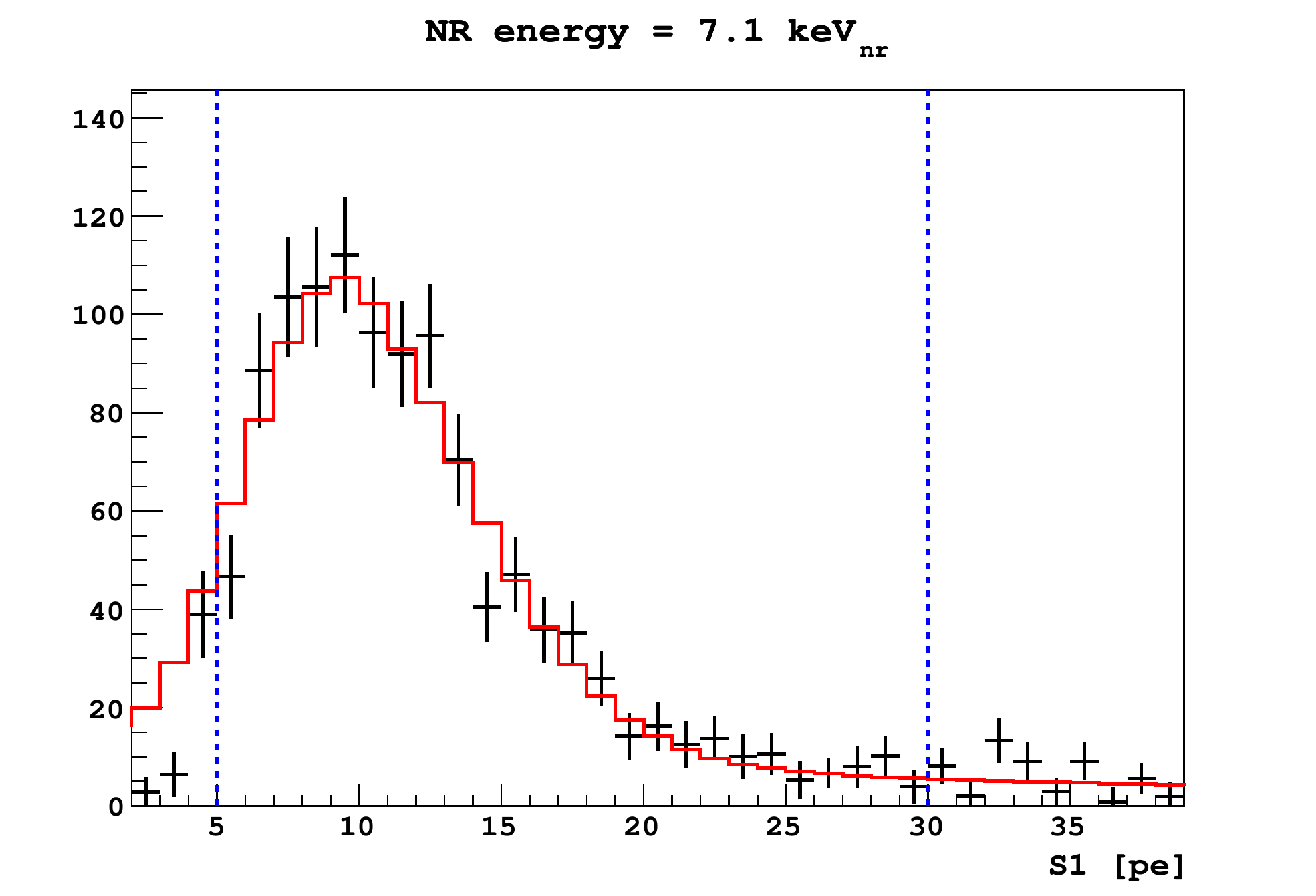}
	\includegraphics[width=2.80in]{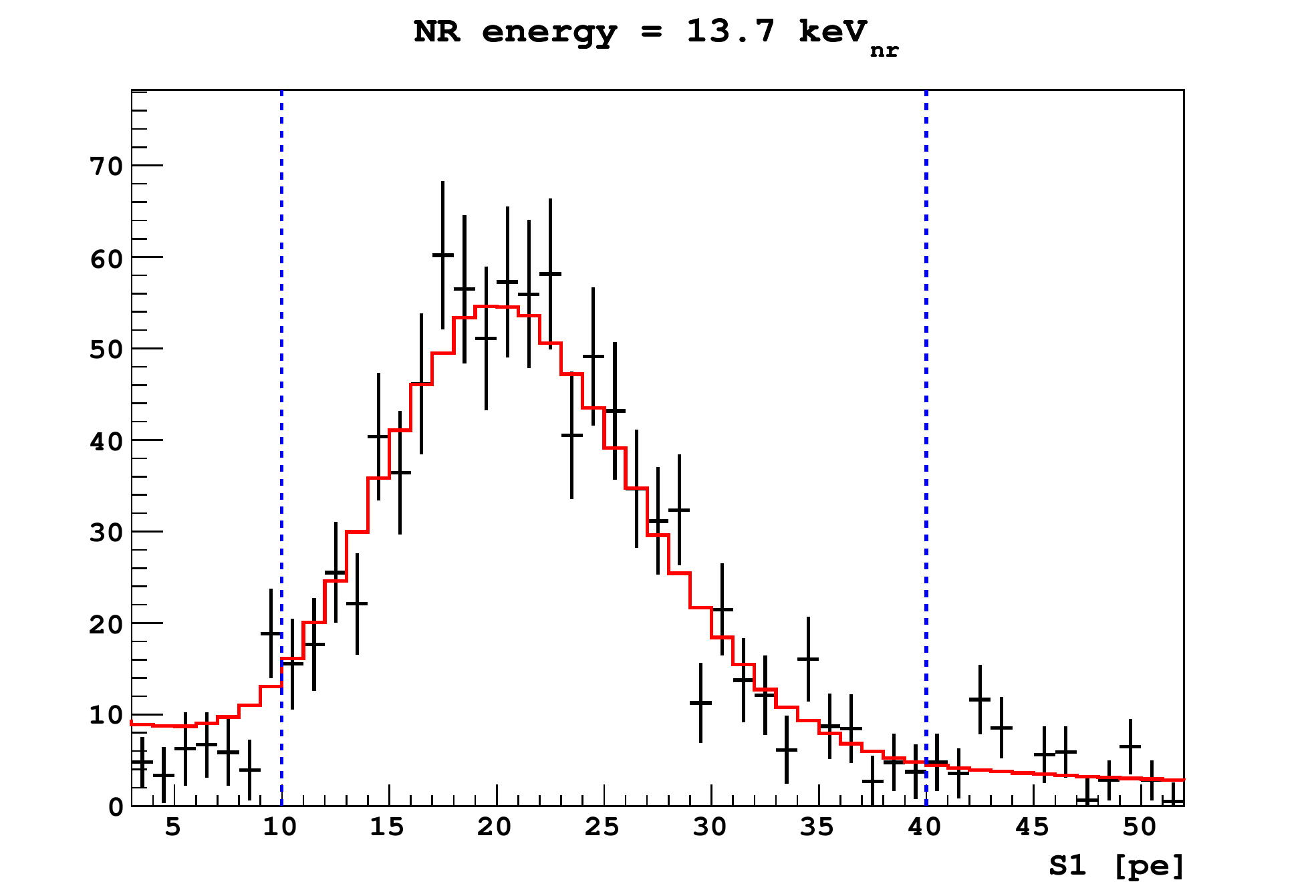}
	\includegraphics[width=2.80in]{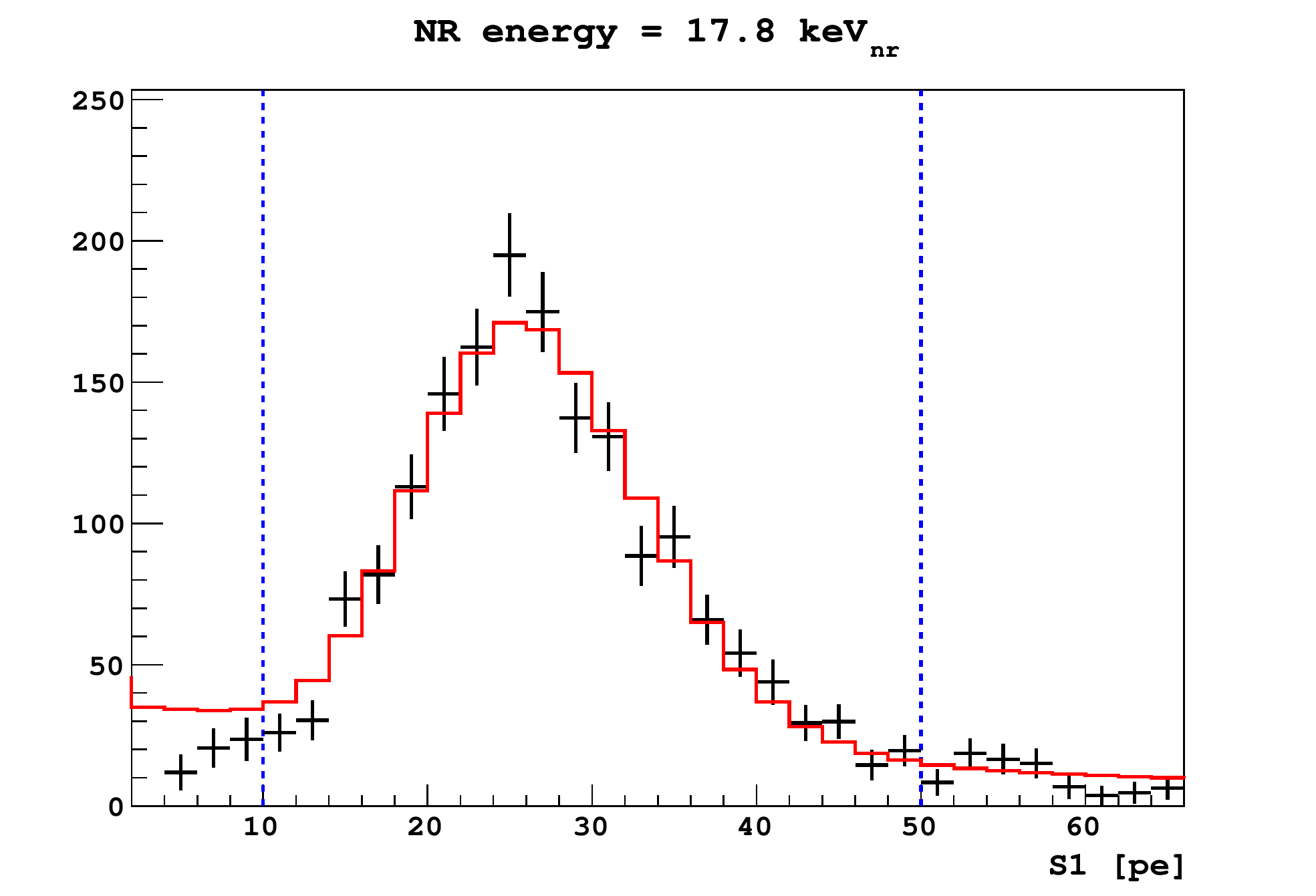}
	\includegraphics[width=2.80in]{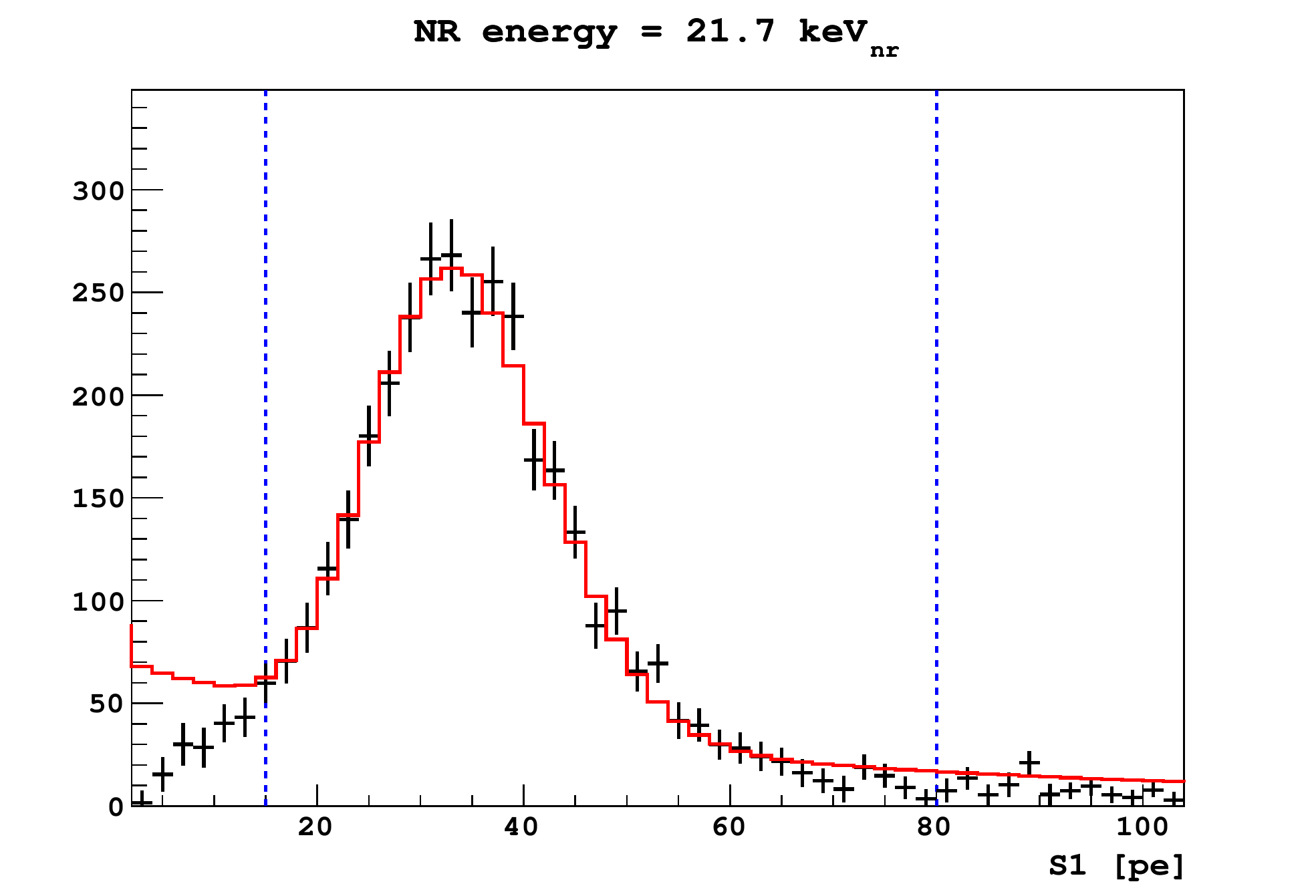}
	\includegraphics[width=2.80in]{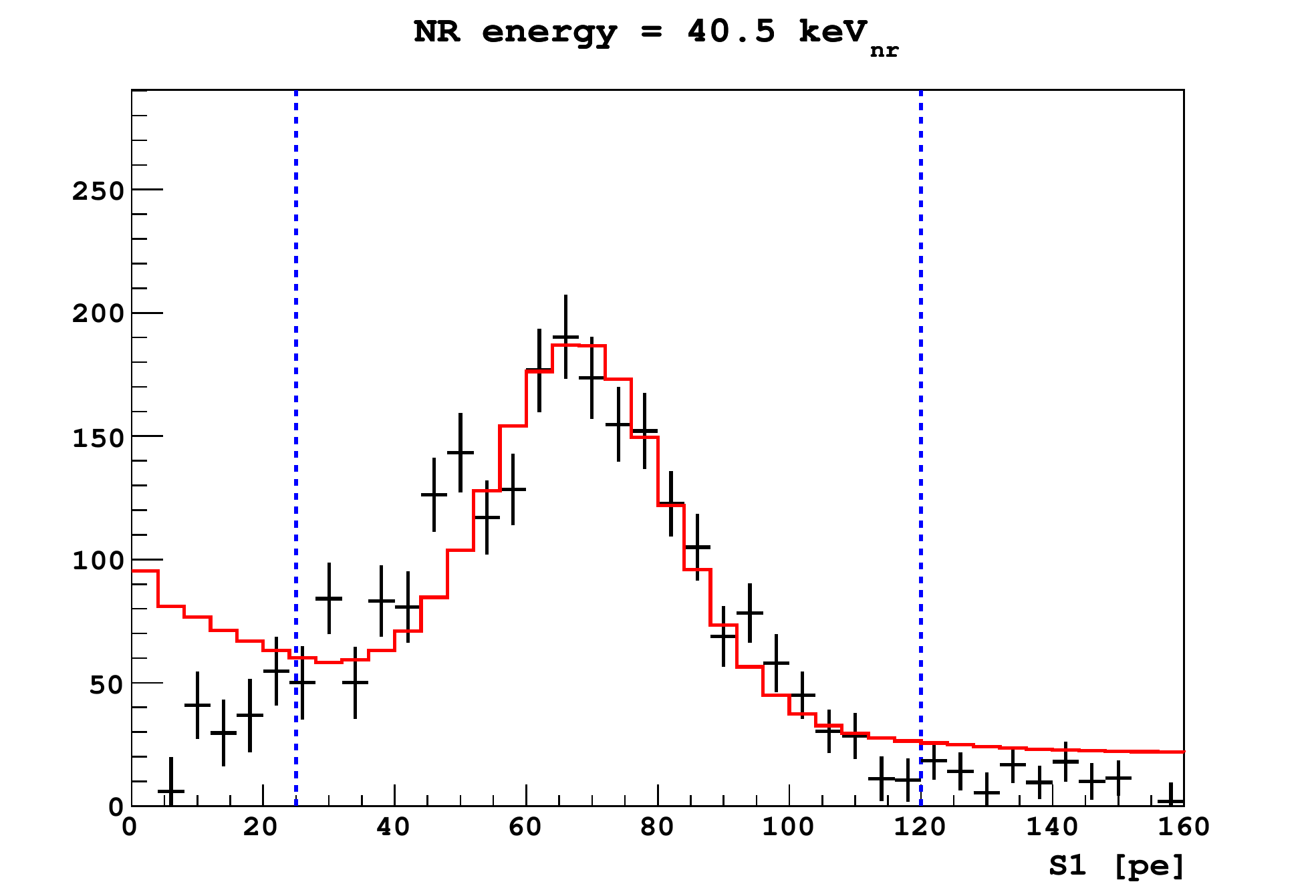}
	\includegraphics[width=2.80in]{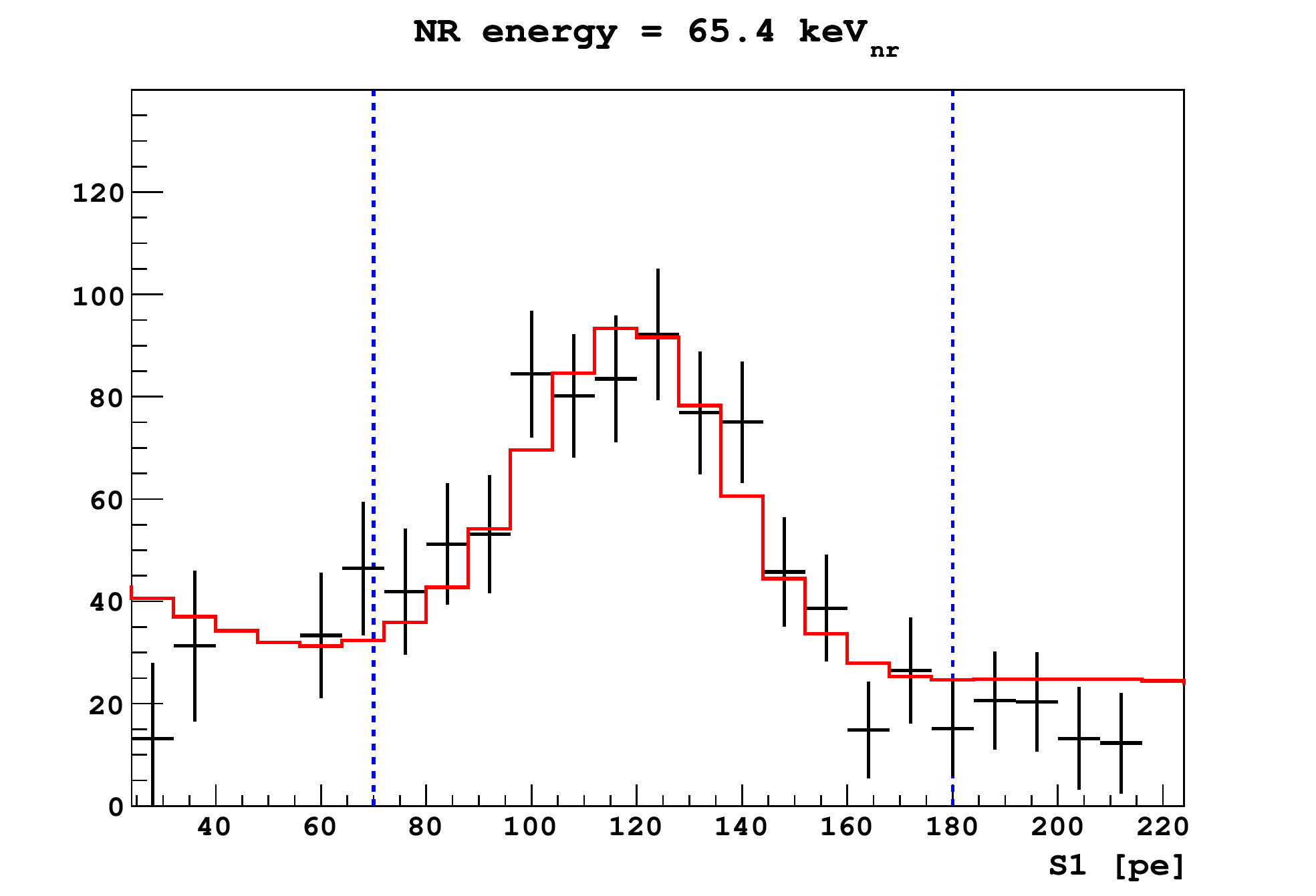}
	\includegraphics[width=2.80in]{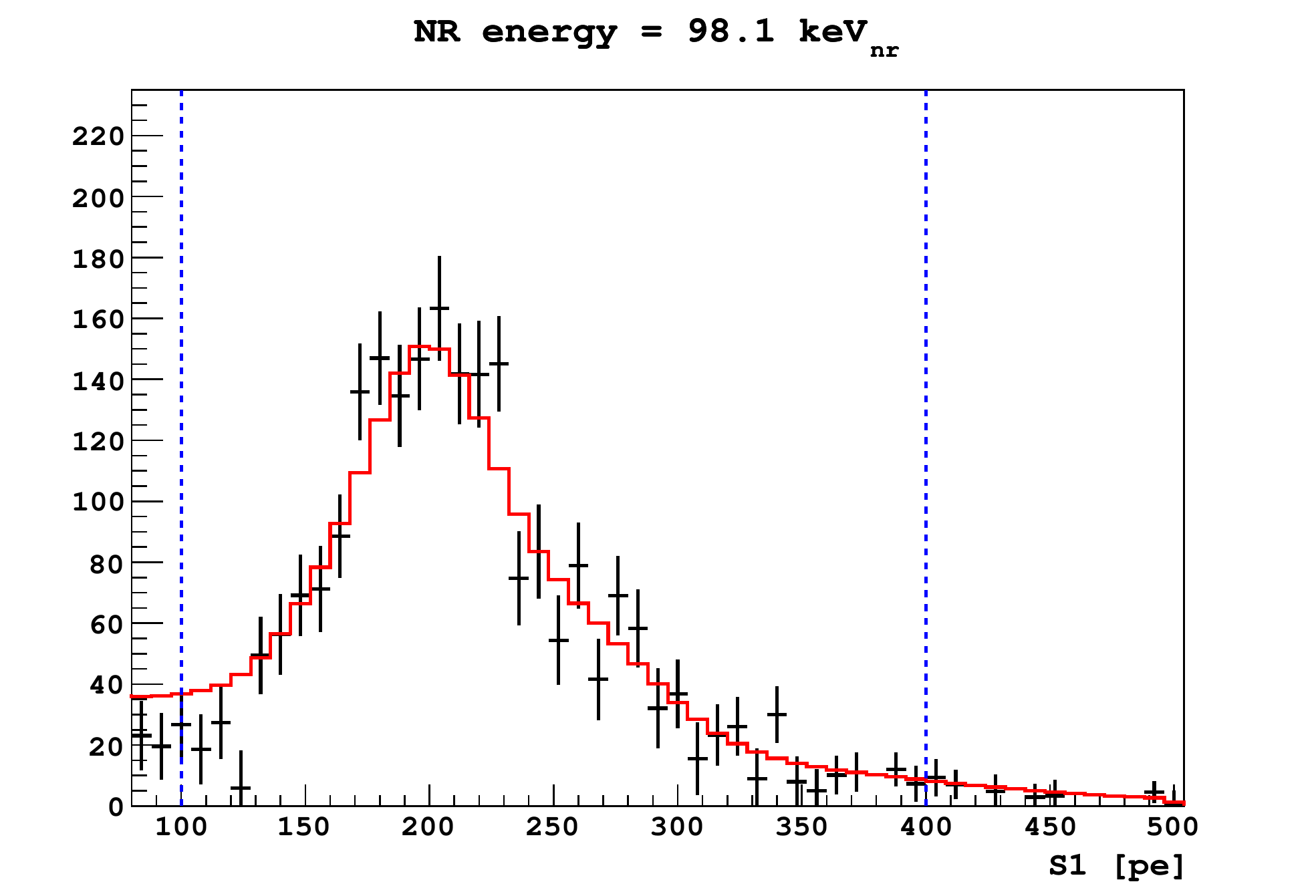}
	\includegraphics[width=2.80in]{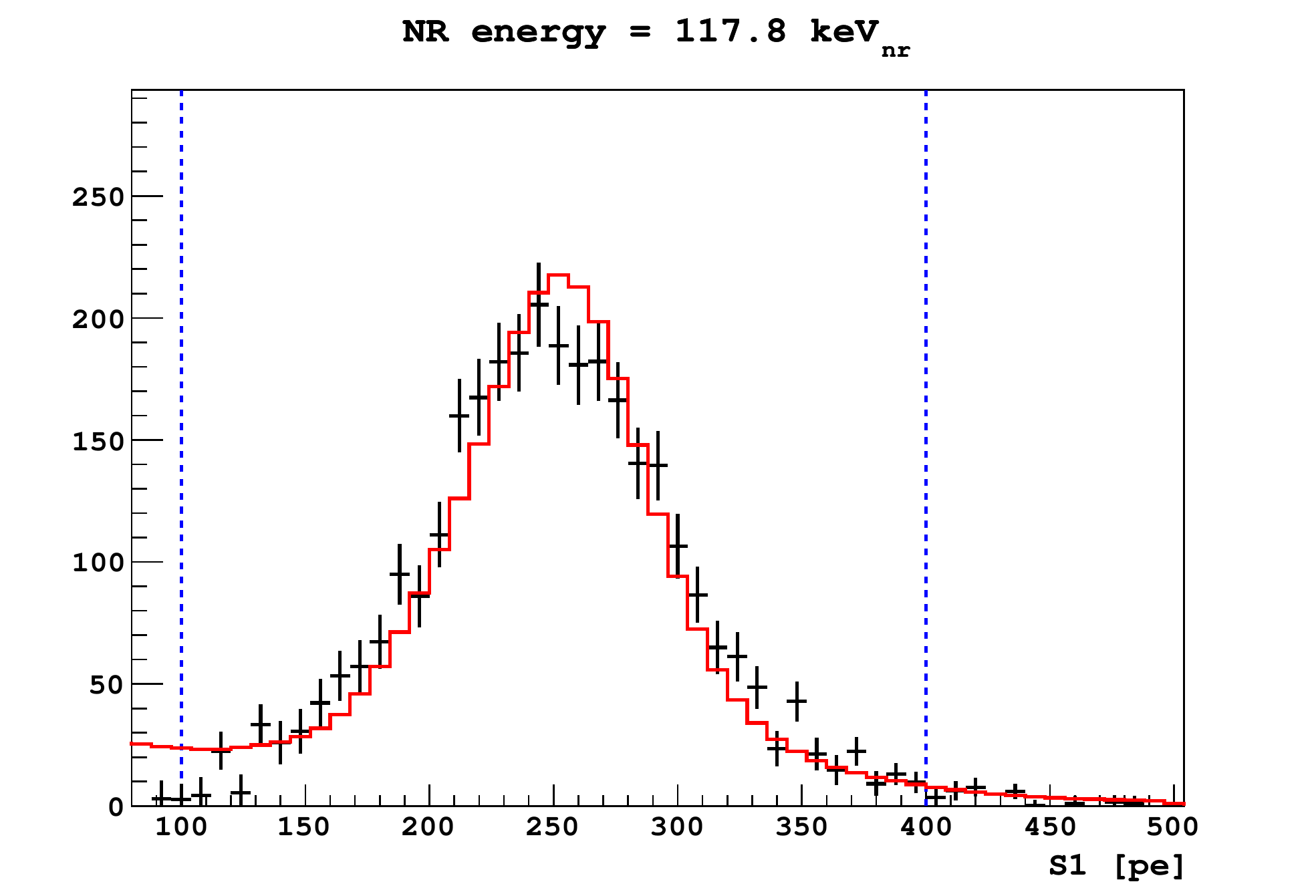}
	\caption{Nuclear recoil data taken with zero electric field, fitted with the Monte Carlo-derived probability density functions for events in coincidence with the A0-7 detectors (red lines).  The vertical dashed lines indicate the fitting range for each spectrum.}
	\label{fig_fit_res_part2}
\end{minipage}
\end{figure*}
\end{centering}

S1 distributions of NR data samples, selected with a coincidence signal from each ND as described in section~\ref{sec:selection}, are independently fit with a probability density function derived from the  Monte Carlo.  The only free parameters in the fit are the normalization factor and \Leff, which acts as a scaling factor of the light yield. 
Results from each fit are shown in figure~\ref{fig_fit_res_part2}.

The sources of systematic error affecting this measurement are listed in table~\ref{tab_sys}. The dominant contributions to the uncertainty on \Leff are the uncertainties on the LY and on the ND positions. The first is evaluated with an analytical propagation of the error on the LY, while the second relies on Monte Carlo simulations where the ND positions were varied according to the uncertainty from the survey in the direction that maximizes the NR energy spread. The survey was done by measuring the distance of each ND from several reference points along the beam direction. An \textit{a posteriori} cross-check was done by overlaying several photographs of the entire setup with the rendering of the geometry in the Monte Carlo using the BLENDER software~\cite{BLENDER}. The TPC, the source position, and the ND support structures were used as reference anchors in the comparison. All ND positions were confirmed within a maximum shift of 4~cm with the exception of A2, which required a shift of (-6,+7,+13)~cm with respect to the survey position\footnote{x is the beam-TPC direction, z is the vertical coordinate, and y is orthogonal to the xz--plane.}. 
The size of the shift is conservatively assumed to be the uncertainty on the A2 position and, when propagated to the NR energy, results in an uncertainty of 5.5\%.
The uncertainty on the NR energy for coincidences with the other NDs ranges from 0.8\% to 2.5\%.   

Other subdominant sources of systematic error relating to the setup geometry and materials are the uncertainties on the \Li energy, with its determination described in section \ref{sec:BeamKinematics}, and the TPC position, known within 1~cm. Their impact on \Leff, quoted in table~\ref{tab_sys}, was evaluated with Monte Carlo simulations. Systematic effects associated with the analysis procedure, such as the trigger efficiency correction, the TOF cuts, histogram binning, energy range of the fit, and background subtractions were investigated by varying the associated parameters. Only the uncertainties on the trigger efficiency and the TOF selection induce a non--negligible systematic error on \Leff and are quoted in table~\ref{tab_sys}. 

\begin{figure*}
\includegraphics[width=0.7\textwidth]{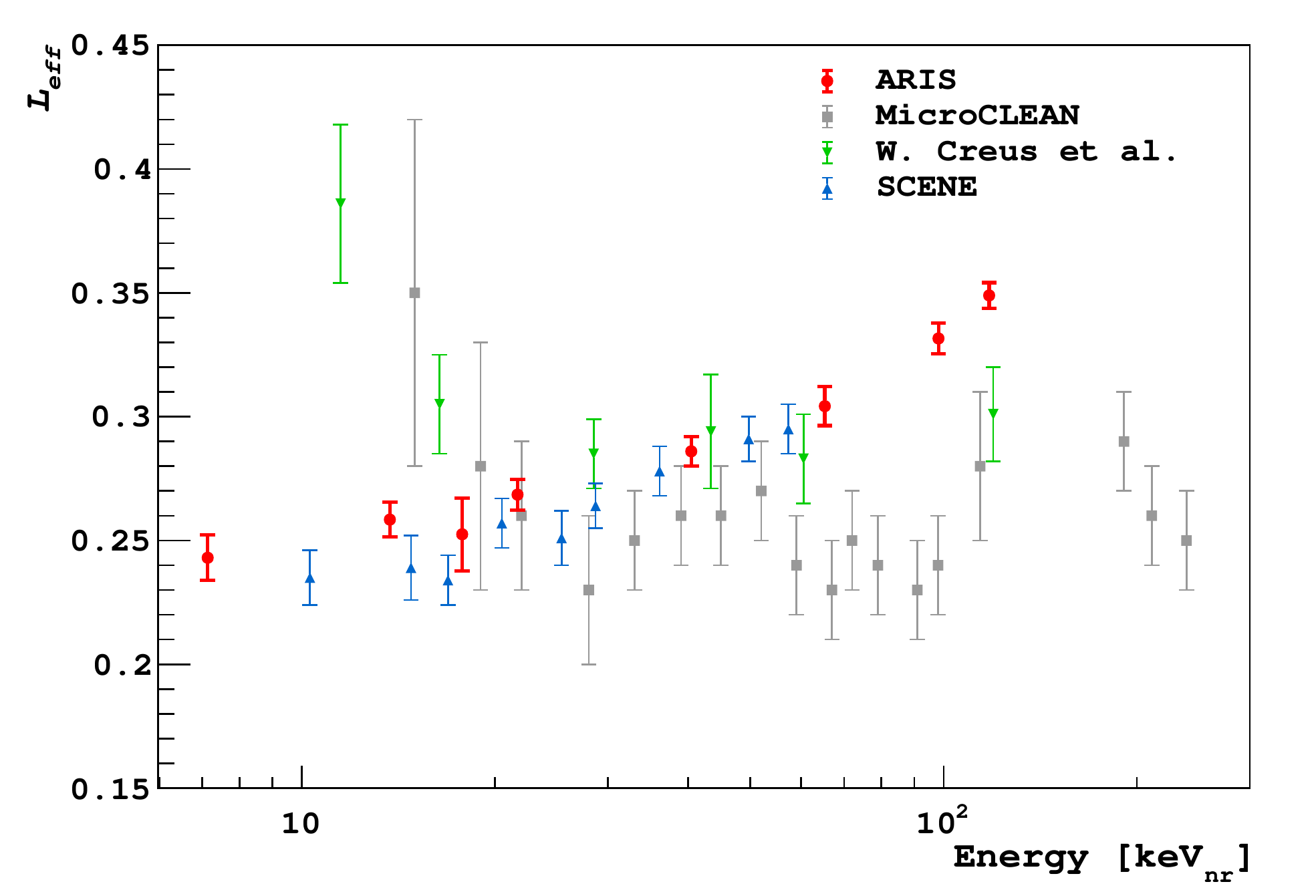}
\caption{\Leff dependence on NR energy as measured by this work and compared with other data sets~\cite{Cao:2015ks, Gastler:2010sc, Creus:2015fqa}.}
\label{fig_quenching}
\end{figure*}

\begin{table*}
  \centering
  \begin{tabular}{lC{1.2cm}C{1.2cm}C{1.2cm}C{1.2cm}C{1.2cm}C{1.2cm}C{1.2cm}C{1.2cm}C{1.2cm}}
    \hline
    \hline
    NR energy [keV] & 7.1 & 13.7 & 17.8 & 21.7 & 40.5 & 65.4 & 98.1 & 117.8\\ 
    \hline
    \Leff & 0.243 & 0.258 & 0.253 & 0.269 & 0.286 & 0.304 & 0.332 & 0.349    \\ 
    \hline
    Light-yield  & 0.002 & 0.002 & 0.002 & 0.002 & 0.002 & 0.002 & 0.003 & 0.003    \\ 
    Beam kinematic  & 0.001 & 0.002 & $o(10^{-3})$ & $o(10^{-3})$ & $o(10^{-3})$ & $o(10^{-3})$ & $o(10^{-3})$ & $o(10^{-3})$    \\ 
    A0--A7 position  & 0.006 & 0.005 & 0.014& 0.005 & 0.004 & 0.004 & 0.003 & 0.003    \\ 
    TPC position  & $o(10^{-3})$  & $o(10^{-3})$  & $o(10^{-3})$  & $o(10^{-3})$  & $o(10^{-3})$  & $o(10^{-3})$  & $o(10^{-3})$  & $o(10^{-3})$     \\ 
    A0--A7 TOF  & $o(10^{-3})$   & $o(10^{-3})$   & 0.001 & 0.001 & $o(10^{-3})$   & 0.002 & 0.001 & 0.001    \\ 
    TPC TOF  & 0.002 & 0.001 & 0.001 & 0.001 & 0.002 & 0.002 & 0.002 & 0.002    \\ 
    Trigger efficiency  & $o(10^{-3})$  & $o(10^{-3})$  & $o(10^{-3})$   & $o(10^{-3})$   & $o(10^{-3})$   & $o(10^{-3})$   & $o(10^{-3})$   & $o(10^{-3})$      \\ 
    \hline
    Total Syst. & 0.007 & 0.006  & 0.014  & 0.006  &  0.005 &  0.005 & 0.005  &   0.005    \\ 
    \hline
    Stat. & 0.005 & 0.004 & 0.003 & 0.002 & 0.003 & 0.006 & 0.004 & 0.002    \\ 
    \hline
    Combined & 0.009 & 0.007 & 0.015 & 0.006 & 0.006 & 0.008 & 0.006 & 0.005    \\ 
    \hline
    Combined relative [\%] & 3.8 & 2.7 & 5.8 & 2.3 & 2.1 & 2.6 & 1.8 & 1.5    \\ 
    \hline
    \hline
  \end{tabular}
  \caption{Measured \Leff for NR events coincident with each ND with the different sources of systematic uncertainties and the statistical uncertainty from the fit quoted.}
  \label{tab_sys}
\end{table*}


This work provides the most precise determination of the \Leff dependence on the NR energy in LAr, as shown in figure~\ref{fig_quenching}, where it is compared with previous measurements~\cite{Cao:2015ks, Gastler:2010sc, Creus:2015fqa} in LAr.  

\begin{figure}
\includegraphics[width=\columnwidth]{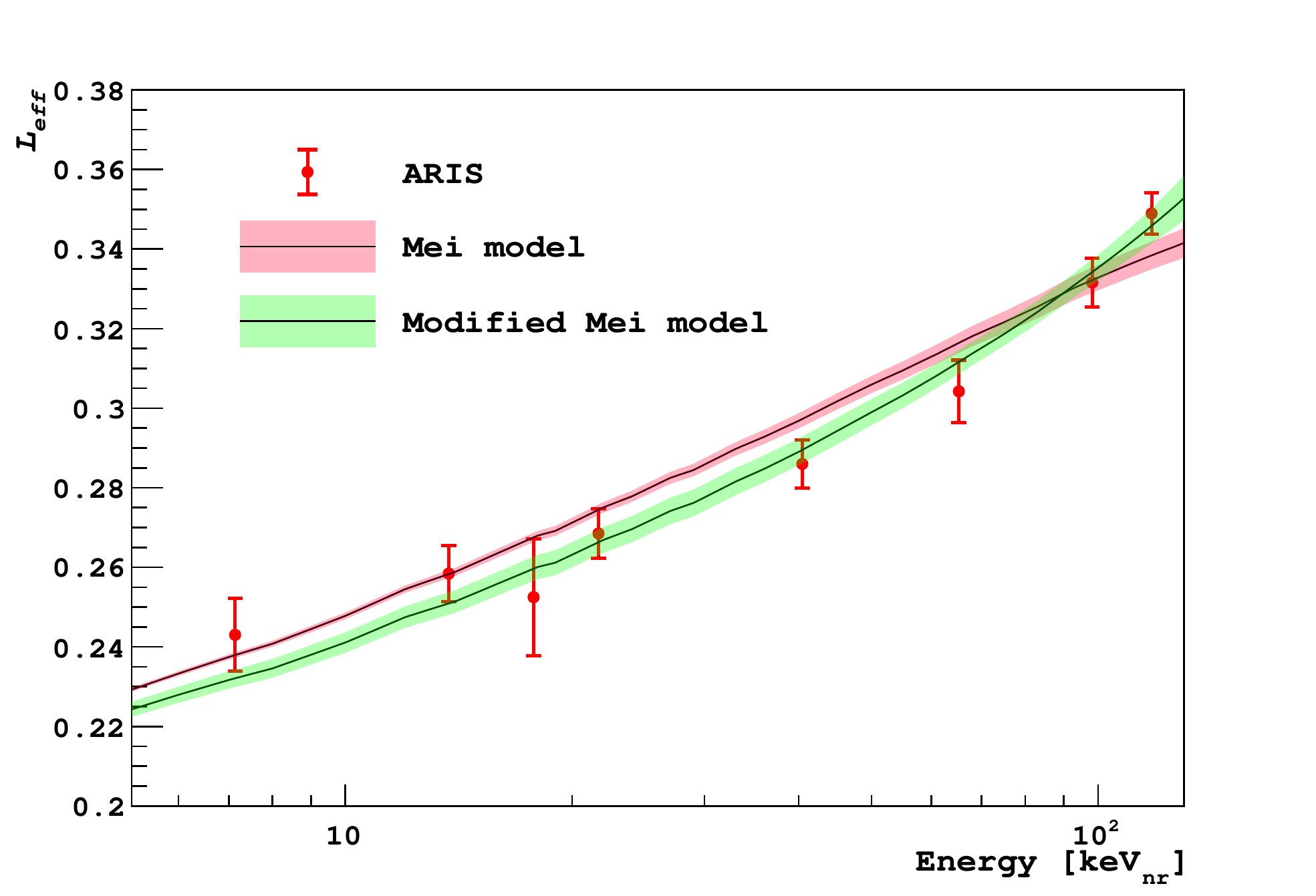}
\caption{$\mathcal{L}_{eff}$ values measured by this work fit with Mei and modified Mei models as described by equations  \ref{eq:mei} and \ref{eq:mei2} in the text.}
\label{fig_quenching_models}
\end{figure}

These results can be compared with the quenching models for LAr proposed by Mei~\cite{Mei:2008ca}, which predicts a quenching factor of 
\begin{equation}
\mathcal{L}_{eff}^{M}= f_n \times \frac{1}{1 + k_B \frac{dE}{dx}},
\label{eq:mei} 
\end{equation} 
where, $f_n$ is  the ionization energy reduction factor   due to losses to the nuclear stopping power, as predicted by Lindhard model~\cite{Lindhard:1963}. The Mei model  derives  $k_B$ = 7.4$\times$10$^{-4}$ MeV$^{-1}$ g cm$^{-2}$ from heavy ion measurements.  Figure \ref{fig_quenching_models} shows, however, this model does not accurately reproduce ARIS data.  The Mei model is disfavored at 2$\sigma$ even using  $k_B$ as a free parameter in a fit. The agreement is recovered by adding a quadratic term, 
\begin{equation}
\mathcal{L}_{eff}^{M*}= f_n \times \frac{1}{1 + k_B \frac{dE}{dx}+ k^{*}_B (\frac{dE}{dx})^2 },
\label{eq:mei2} 
\end{equation}
as in the extended version of the Birks'  formula for organic scintillators \cite{SMITH1968157}. In this way, the model is compatible with the data with a p-value of 0.79 as shown in figure~\ref{fig_quenching_models}, and the   best fit parameters are \mbox{$k_B$ = (5.2$\pm$0.6)$\times$10$^{-4}$ MeV$^{-1}$ g cm$^{-2}$} and \mbox{$k^{*}_B$ = (-2.0$\pm$0.7)$\times$10$^{-7}$ MeV$^{-2}$ g$^2$ cm$^{-4}$}.  This result is in agreement with the best fit of the modified Mei model to DarkSide-50 data which yields a value of \mbox{$k_B$ = (4.66$^{+0.86}_{-0.94}$)$\times$10$^{-4}$ MeV$^{-1}$ g cm$^{-2}$}~\cite{Agnes:2017grb}.



\section{S1 response versus electric field}
\label{sec:recombination}

In addition to the null field data set, data were acquired  at 50, 100, 200,  and 500 V/cm drift fields in triple coincidence mode.  The presence of an electric field in the active volume increases the probability  that ionization electrons   escape the electron-ion cloud, reducing the recombination effect. 

Any energy deposit  in LAr   produces an average number of quanta ($N_q$), either  excitons or ion-electron pairs, corresponding to   

\begin{equation}
N_q =  N_i + N_{ex} =  \mathcal{L}_{eff} \times { E_{dep} \over  W } .
\end{equation}

where $W$=19.5~eV \cite{Doke:1988kw} is the effective work function,  $N_{ex}$  and    $N_{i}$  the numbers of excitons and ions, respectively, and  where $\mathcal{L}_{eff}$ is assumed to be 1 for \ERs. S1 can be expressed as function of  $\alpha$,  the   $N_{ex}$/$N_{i}$ ratio: 

\begin{equation}
S1    = \epsilon_1~(\alpha  + R) \times N_{i}
\label{eq:s1reco}
\end{equation}
where $\epsilon_1$ is the light collection efficiency of the detector, and $R$ the electron-ion recombination probability. The value of $\alpha$ is equal to 0.21  for ERs,  and to 1 for NRs~\cite{Doke:1988kw}.

In ARIS, the recombination dependences on electron recoil equivalent energy  (E$_{ee}$)\footnote{In case of NRs, E$_{ee}$ = \Leff(E$_{nr}$)$\times$E$_{nr}$.  }  and field (F) are studied with respect to the observable:

\begin{equation}
{ S1 \over S1_{0} } = { \alpha + R (E_{ee}, F) \over 1 + \alpha},
\label{eqn:zvariable}
\end{equation}

where S1$_0$ is the scintillation response at null field. 

\begin{figure}[t]
\includegraphics[width=\columnwidth]{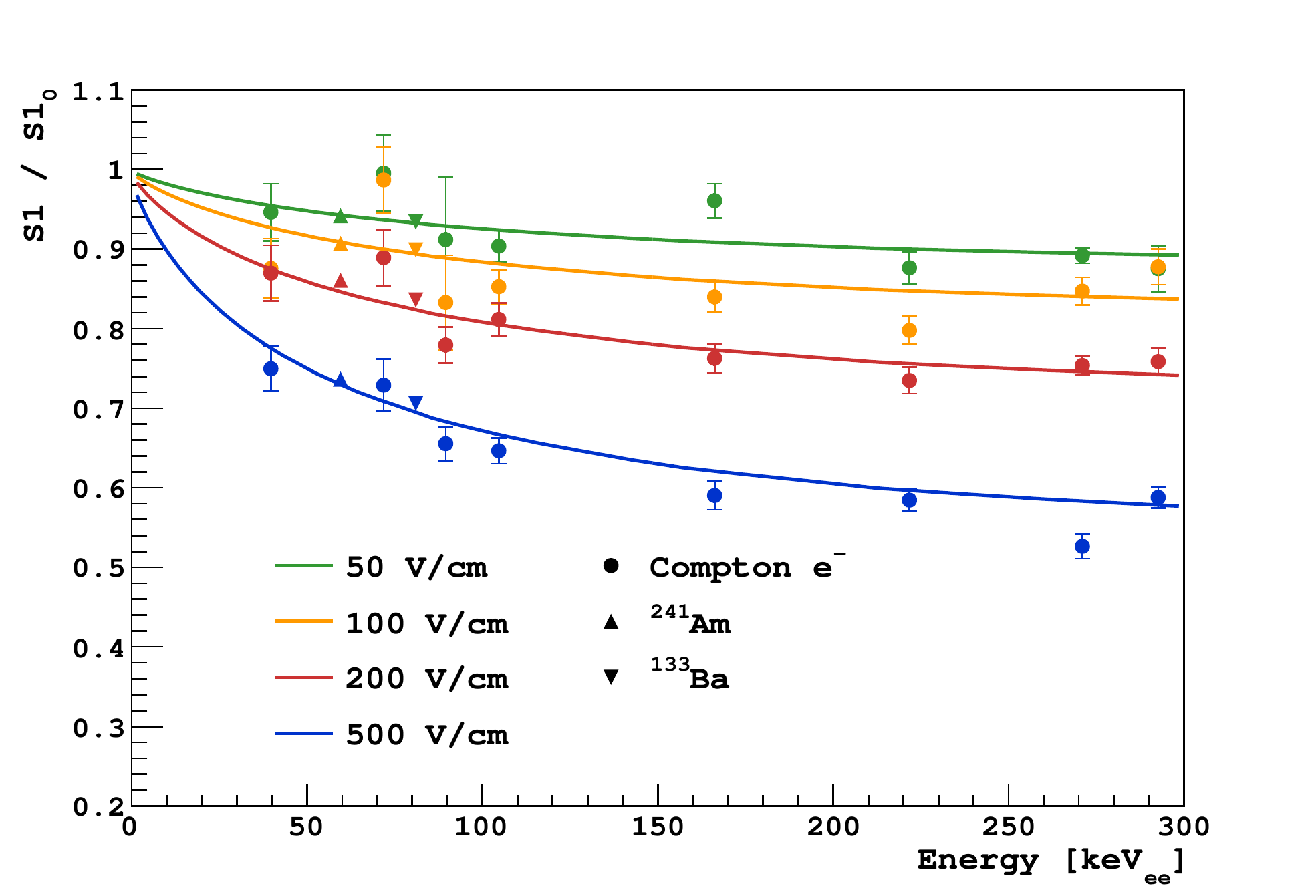}
\caption{Field induced quenching of S1 for \ERs\ at different drift fields, fit with the Doke-Birks model.}
\label{fig:recombination_ER}
\end{figure}

Equation \ref{eqn:zvariable}  is expected to reproduce ARIS data in both ER and NR modes, by accordingly changing the value of $\alpha$, if the   recombination probability $R(E,F)$ is correctly modelled. 
The S1/S1$_0$ ratio extracted from the data is compared with three recombination models:
Thomas-Imel~\cite{Thomas:1987zz},  Doke-Birks~\cite{Doke:1988kw},  and  PARIS model~\cite{Agnes:2017grb}.  The first is an extension of the Jaffe ``box'' theory \cite{ANDP:ANDP19133471205} and was demonstrated to be accurate in the ``short track'' regime, such as NRs or low energy ERs. The Doke-Birks model is empirical and expected to reproduce data at higher energies. PARIS was tuned on DarkSide-50 data at 200 V/cm only, but was demonstrated to work from $\sim$3~keV up to $\sim$550~keV.

The Doke-Birks model parametrizes $R$ as the following:

\begin{equation}
R = \frac{A~dE/dx}{1+B~dE/dx} + C,
\end{equation}
where B = A/(1-C) and $dE/dx$ is the energy loss by electrons in LAr. We introduce a  dependence on the electric field, $F$, by defining 

\begin{equation}
C = C'\,e^{-D\times F}. 
\end{equation}

\begin{figure} [!t]
\includegraphics[width=\columnwidth]{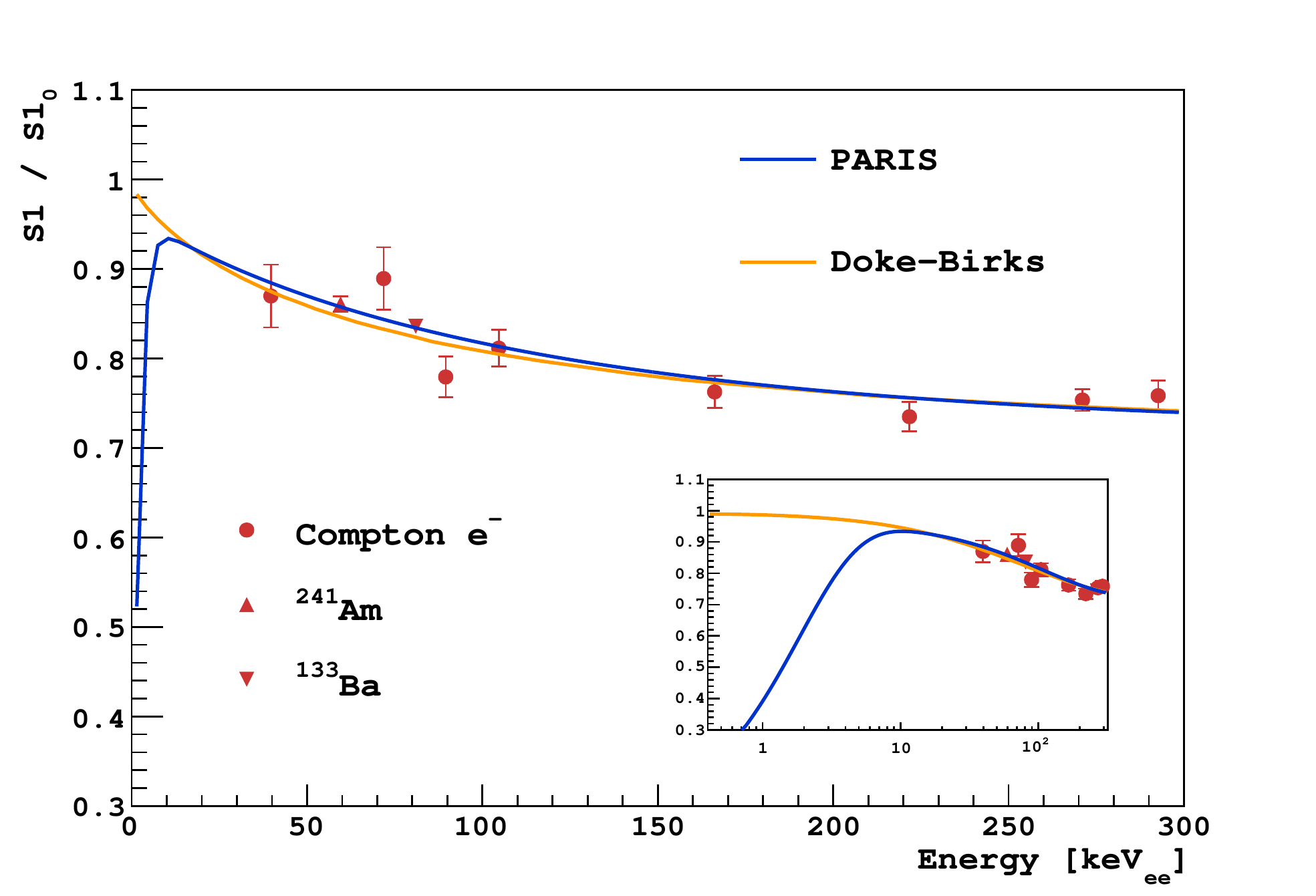}
\caption{Field induced quenching of S1 for \ERs\ at 200 V/cm compared with the PARIS model and the fit of the Doke-Birks model.  The inset shows the same data with the x-axis represented in a log-scale.}
\label{fig:recombination_PARIS}
\end{figure}

ARIS data in ER mode were simultaneously fit with this electric field-modified version of  Doke-Birks  in the [40,300]~keV$_{ee}$ range, with the results shown in figure \ref{fig:recombination_ER}. The parameters returned by the fit are   A=(2.5$\pm$0.2)$\times$10$^{-3}$ cm/MeV, C'=0.77$\pm$0.01, and D=(3.5$\pm$0.3)$\times$10$^{-3}$ cm/V. With these parameters, the model is able to reproduce ER data with energy from 40 keV at any field. However, while the Doke-Birks   recombination tends to 1  at lower energies, observations from the DarkSide-50 data demonstrate that it should decrease \cite{Agnes:2017grb}. The PARIS model, which was designed to solve this issue, does not require any tuning of the parameters and accurately matches the data, as shown in figure~\ref{fig:recombination_PARIS}.     The difference between Doke-Birks and PARIS models appears for energies below 10~keV$_{ee}$. 

NR data, converted in ER equivalent energy by means of the \Leff measured in section \ref{sec:quenching}, are fit with the  Thomas-Imel model, in which the recombination probability is given by  

\begin{equation}
R = 1 - \frac{ln(1 + \xi)}{\xi} ,
\end{equation}

where 
\begin{equation}
\xi = C_{box}\frac{N_i}{F^\beta}.
\end{equation}

$C_{box}$ is a constant depending on  the mean ionization electron velocity $v$ and on the size of the ideal box containing the electron-ion cloud.

\begin{figure}
\includegraphics[width=\columnwidth]{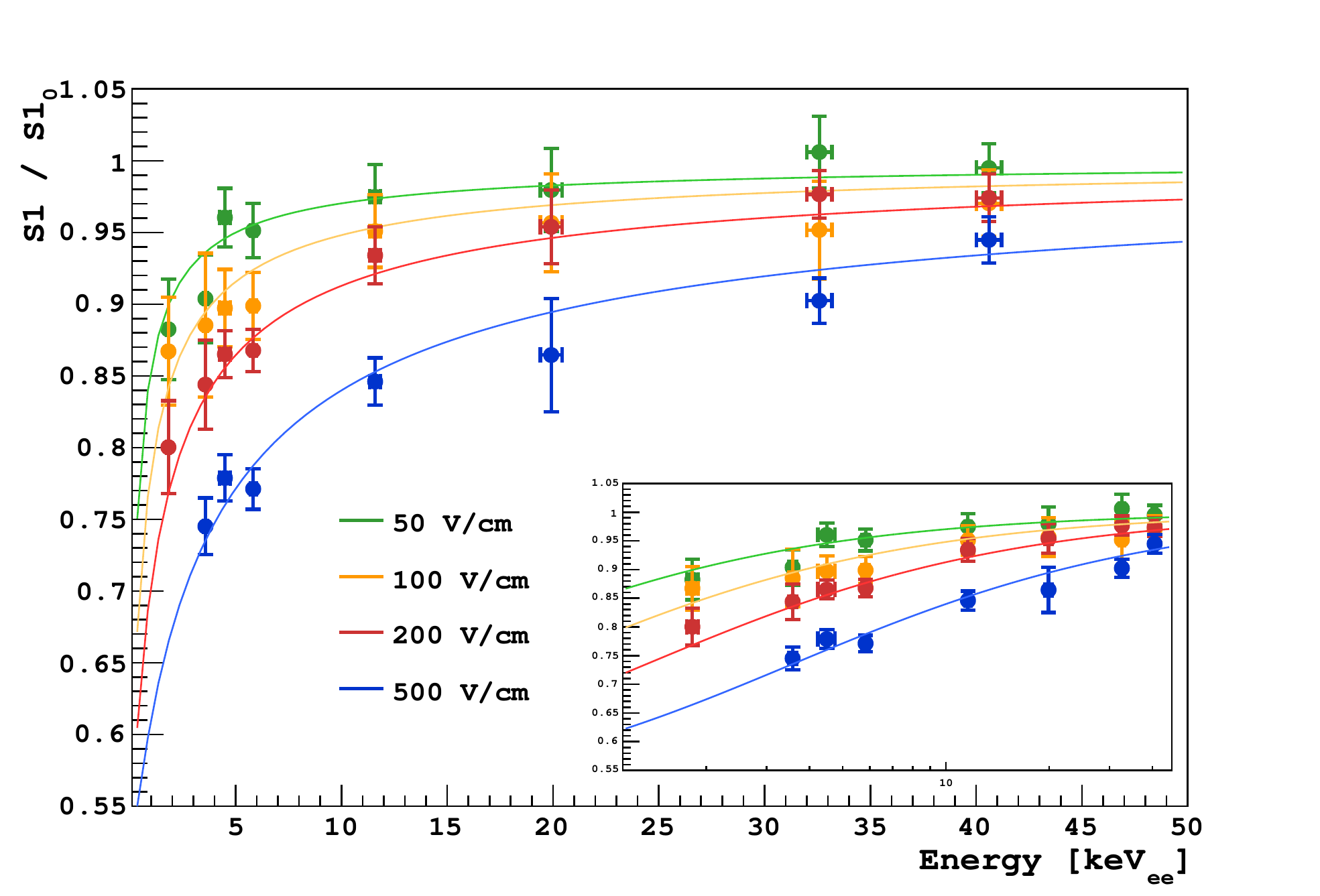}
\caption{Field induced quenching of S1 for \NRs\ for different drift fields fit with the Thomas-Imel model. The systematic uncertainties are included in the error bars.  The inset shows the same data with the x-axis represented in a log-scale.}
\label{fig:recombination_NR_field}
\end{figure}

\begin{figure}
\includegraphics[width=\columnwidth]{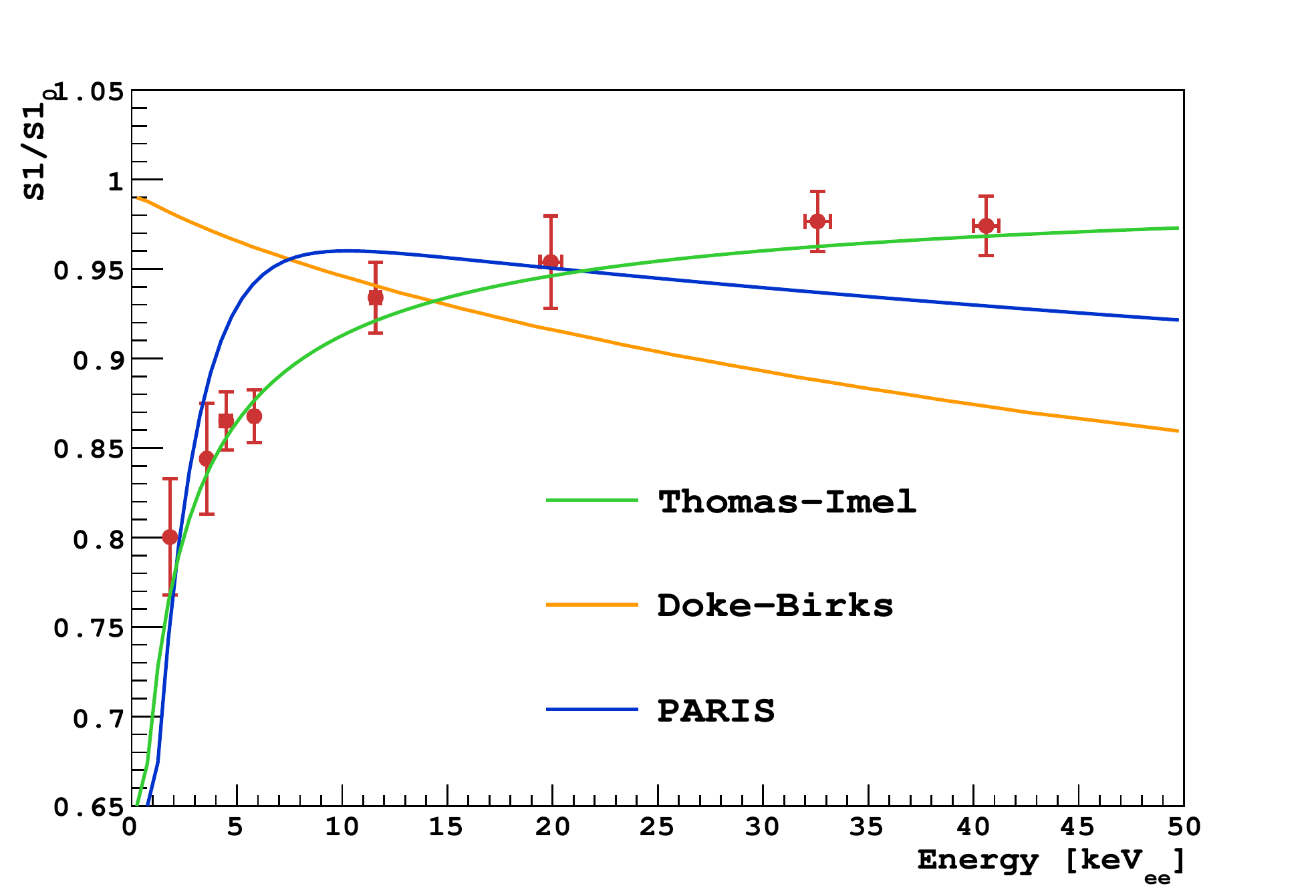}
\caption{Field induced quenching of S1 for \NRs\ at 200~V/cm compared to model predictions from Thomas-Imel, tuned on the NR data set, and Doke-Birks and PARIS models, tuned on ERs, assuming $\alpha$=1. }
\label{fig:recombination_NR_g4ds}
\end{figure}

Figure \ref{fig:recombination_NR_field} shows the S1/S1$_0$ ratio, at different fields, for NRs, fit with the Thomas-Imel model. The fit returns   $\beta$=1.07$\pm$0.09, in good agreement with the Thomas-Imel prediction of $\beta$=1, and $C_{box}$=18.5$\pm$9.7.  The resulting Thomas-Imel model for NRs is compared with Doke-Birks and PARIS under the paradigm that, with a fixed recombination probability, models should be able to describe both ER and NR data sets by changing the scintillation-to-ionization ratio from $\alpha$=0.21 (ER) to $\alpha$=1 (NR). This paradigm is disproved by the comparison between models and the NR data set at 200 V/cm, shown in figure~\ref{fig:recombination_NR_g4ds}, where Doke-Birks and PARIS predictions are rejected at more than 5$\sigma$.  The Doke-Birks and PARIS models are not recovered in NR mode, even by changing the  $\alpha$ value. 

An overall model requires then two separate recombination probabilities in order to describe both ERs and NRs. In  the range of dark matter searches in LAr ($<$60~keV$_{ee}$),  the tuned Thomas-Imel model was demonstrated to correctly describe scintillation response to NRs, while PARIS is confirmed as a good modeling for ERs, if operating at 200~V/cm. Doke-Birks provides a good description of ERs at different fields, but almost outside the range  of interest ($>$40~keV$_{ee}$) for dark matter searches. 

\begin{figure}
\includegraphics[width=\columnwidth]{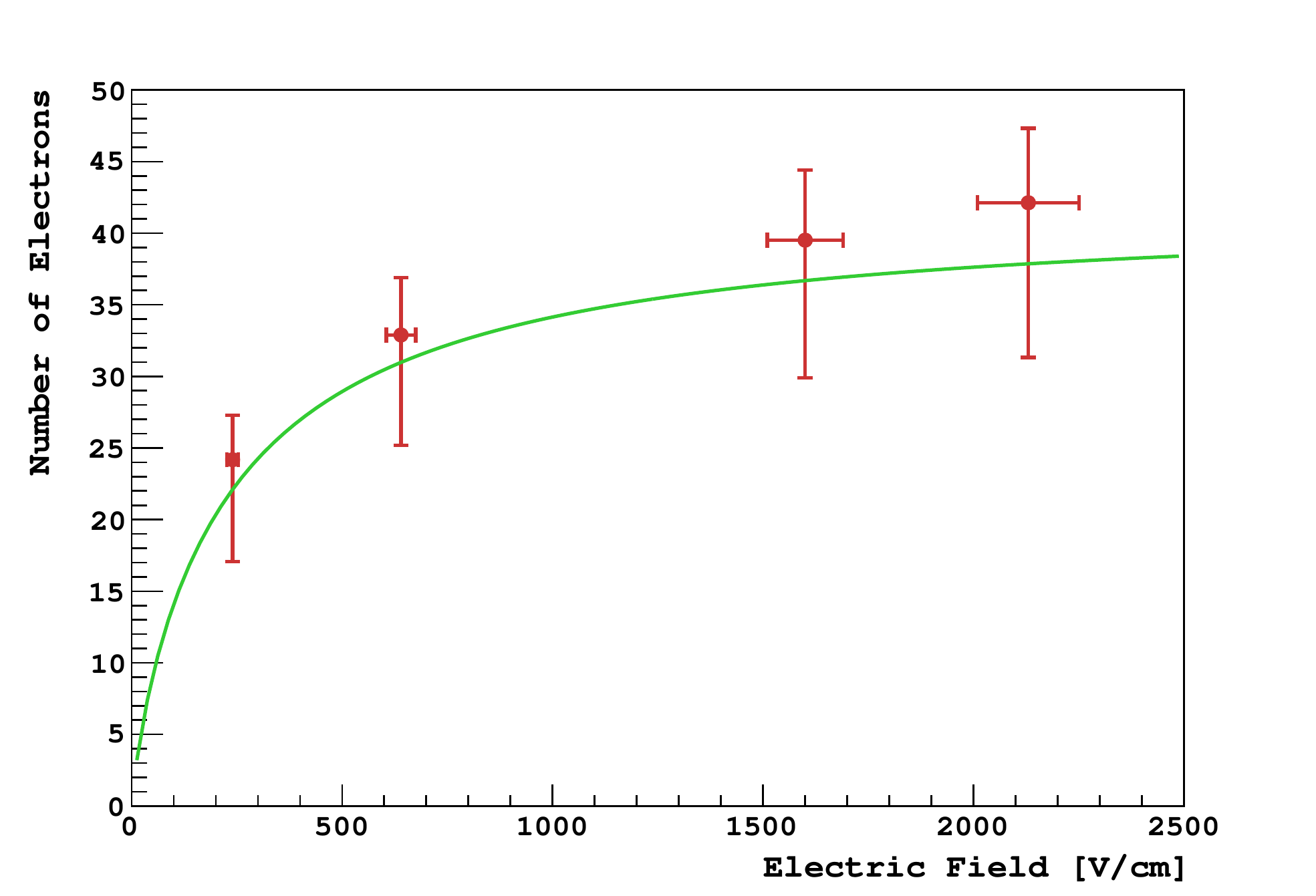}
\caption{Comparison of the S2 signal, expressed in number of ionization electrons,  between the \textit{Joshi et al.} data set at 6.7~keV$_{nr}$ and the Thomas-Imel model prediction, as a function of the drift field.}
\label{fig:joshi}
\end{figure}

As a final check, the tuned Thomas-Imel model was used for predicting the number of  ionization electrons escaping the recombination

\begin{equation}
N_e =  \mathcal{L}_{eff} \times { E_{dep} \over  W } \frac{1-R(E_{ee},F)}{1+\alpha}
\end{equation}

measured by Joshi et al. \cite{Joshi:2014fna}, as function of the drift field ($F$), for 6.7~keV$_{nr}$ NRs, assuming the \Leff measured in this work. Figure~\ref{fig:joshi} shows excellent agreement, suggesting that, apart from \Leff, no extra-quenching factor affects  S2, which can be essentially modeled as  complementary to S1, under the assumption that the excitation to ionization ratio $\alpha$ is equal to 1 for NRs.

\section{Conclusions}
\label{sec:conclusions}

The scintillation yield of nuclear recoils relative to electronic recoils between $\sim$7~keV$_{nr}$ and $\sim$120~keV$_{nr}$ has been measured in LAr using a highly collimated and quasi-monoenergetic neutron source. This work presents the most precise measurement and lowest energy probe of \Leff, the nuclear recoil scintillation efficiency  in LAr. 

In addition,  Compton electrons induced by $\gamma$s from \Lim de-excitation, in coincidence  with the neutron beam, are used to measure the relative scintillation LY  as a function of energy and drift field along with $\gamma$s from calibration sources.  At null field, the LY  was measured to be constant within 1.6\% in the [40, 511]~keV$_{ee}$ range, the most stringent test of the linearity of the LAr response. Furthermore, no differences were observed in the light response   between single- and multi-scatter ER events. 

In presence of an electric field, three models (Thomas-Imel, Doke-Birks, and PARIS) were compared to the NR and ER data sets. The Thomas-Imel electron-ion recombination probability function, properly tuned on these data, provides a good description of the response to NRs at different fields, while PARIS is confirmed as a good model for ERs at the DarkSide-50 operation drift field of 200 V/cm.  The   Doke-Birks recombination probability models the response to ERs at different fields, but only above 40~keV$_{ee}$, in the upper range  of interest for dark matter searches. 

Finally, a comparison of the ionization signal between the tuned Thomas-Imel model and an independent NR data set at 6.7~keV$_{nr}$ suggests  that no extra quenching factors are required to predict  the number of ionization electrons.

In conclusion, this work provides a fully comprehensive model of the  LAr response in the range of interest for dark matter searches through measurement of the \Leff parameter as a function of NR energy, and by properly tuning the parametrization of the electron-ion recombination  probabilities for  ERs and NRs.

Recent analyses of DarkSide-50 have  extended  by up to an order of magnitude the exclusion region for WIMP-nucleus interactions  in the WIMP mass range below 6 GeV/c$^2$ \cite{Agnes:2018ves},  and slightly improved limits for WIMP-electron interactions, assuming a heavy mediator \cite{Agnes:2018oej}. To achieve these results, the DarkSide-50 collaboration has benefitted from the ARIS results by better constraining the response of nuclear recoils in LAr at both field-on and field-off configurations.  The linearity of the electron recoil scintillation response measured by ARIS has allowed DarkSide-50 to derive the spectral shape  of forbidden $^{39}$Ar $\beta$ decay, an important cosmogenic background intrinsic to LAr. The ARIS results have then impacted both the analyses by improving signal and  background models.

\begin{acknowledgments}
The ARIS Collaboration would like to thank IPNO and its staff for invaluable technical and logistical support. We also thank Stephen Pordes for providing  electronic boards.  This report is based upon work supported by the US NSF (Grants PHY-1242585, PHY-1242611, PHY-1314501, PHY-1314483, PHY-1314507, and PHY-1455351) and France-Berkeley Fund (2016-0053). We acknowledge the financial support from the UnivEarthS Labex program of Sorbonne Paris Cit\'e (ANR-10-LABX-0023 and ANR-11-IDEX-0005-02).  

\end{acknowledgments}
\bibliographystyle{ds}
\bibliography{ds}
\end{document}